\documentstyle[emulateapj,11pt,psfig,lscape]{article}

\makeatletter

\newenvironment{inlinefigure}{%
\def\@captype{figure}%
\noindent\begin{minipage}{0.999\linewidth}\begin{center}}
{\end{center}\end{minipage}\smallskip}
\makeatother

\begin{document}
\title{A large sample of spectroscopic redshifts in the ACS-GOODS 
region of the HDF-N}
\author{L.~L.\ Cowie,\altaffilmark{1,2} 
A.~J.\ Barger,\altaffilmark{1,3,4,2} 
E.~M.\ Hu,\altaffilmark{1,2} 
P.~Capak\altaffilmark{1,2} 
A.~Songaila\altaffilmark{1,2} 
}

\altaffiltext{1}{Visiting Astronomer, W. M. Keck Observatory, which is jointly
  operated by the California Institute of Technology, the University of
  California, and the National Aeronautics and Space Administration.}
\altaffiltext{2}{Institute for Astronomy, University of Hawaii,
  2680 Woodlawn Drive, Honolulu, HI 96822; cowie@ifa.hawaii.edu; 
  barger@ifa.hawaii.edu; hu@ifa.hawaii.edu; capak@ifa.hawaii.edu}
\altaffiltext{3}{Department of Astronomy, University of
Wisconsin-Madison, 475 North Charter Street, Madison, WI 53706.}
\altaffiltext{4}{Department of Physics and Astronomy,
University of Hawaii, 2505 Correa Road, Honolulu, HI 96822.}

\lefthead{Cowie et al.\/}

\slugcomment{Submitted to the Astronomical Journal}

\begin{abstract}

We report the results of an extensive spectroscopic survey of
galaxies in the roughly 160~arcmin$^2$ ACS-GOODS region surrounding
the HDF-N. We have identified 787 galaxies or stars with $z'<24$,
$R<24.5$, or $B<25$ lying in the region. The spectra were obtained
with either the DEIMOS or LRIS spectrographs on the Keck 10~m
telescopes. The results are compared with photometric redshift
estimates and with redshifts from the literature, as well 
as with the redshifts of a parallel effort led by a group 
at Keck. Our sample, when combined with the literature data, 
provides identifications for 1180 sources.  

We use our results to determine the redshift distributions
with magnitude, to analyze the rest-frame color distributions 
with redshift and spectral type, and to investigate
the dependence of the X-ray galaxy properties on the 
local galaxy density in the redshift interval $z=0-1.5$. 
We find the rather surprising result that the galaxy X-ray properties 
are not strongly dependent on the local galaxy density for galaxies 
in the same luminosity range. 
\end{abstract}


\section{Introduction}
\label{secintro}

The Hubble Deep Field-North (HDF-N; \markcite{williams96}Williams et al.\ 1996;
\markcite{ferguson00}Ferguson, Dickinson, \& Williams 2000) 
remains the single most intensively studied region on the sky. 
In addition to the defining {\it HST} WFPC2 observations of the 
central 5.3~arcmin$^2$ area (the HDF-N proper), the field was also the 
target of the deepest X-ray exposure to date, the 2~Ms {\it Chandra} 
Deep Field-North (CDF-N) observation 
(\markcite{alex03}Alexander et al.\ 2003; 
\markcite{barger03}Barger et al.\ 2003), which covers about 460~arcmin$^2$. 
Even wider areas have been covered with ultradeep ground-based optical 
and near-infrared imaging (\markcite{capak03}Capak et al.\ 2003 and 
references therein) and with very deep 20~cm radio observations with 
the VLA (\markcite{richards00}Richards 2000).

Most recently, a roughly 160~arcmin$^2$ region, covering the
HDF-N proper and lying mostly within the CDF-N, was observed with the 
ACS camera on {\it HST} in four passbands (F435W, F606W, F775W, F850LP)
as part of the GOODS {\it HST} Treasury program (hereafter, ACS-GOODS; 
\markcite{gia03}Giavalisco et al.\ 2003). 
This region will also be the target of the deepest 
{\it SIRTF} observations, as part of the GOODS 
{\it SIRTF} Legacy program (M. Dickinson, PI).
While not quite as deep as the HDF-N proper, the ACS-GOODS survey 
provides a very deep, high spatial resolution image of a moderately 
large area of sky. 

While photometric redshift surveys can cover large areas---avoiding the 
problems of cosmic variance, which can be significant in 
areas the size of the ACS-GOODS region---and are useful for determining 
the evolution of the luminosity function and 
the universal star formation history
(see \markcite{capak04}Capak et al.\ 2004 for a photometric redshift 
analysis of the 600~arcmin$^2$ Suprime-Cam field around the HDF-N), 
spectroscopic redshift surveys are needed to determine
the evolution of clustering 
and the environmental dependences of galaxy properties, 
such as their X-ray, submillimeter, and radio emission.
Spectroscopic redshifts are also
required for determining the distribution of spectral types,
which may be combined with galaxy morphologies from
the ACS-GOODS observations, and for determining line luminosities,
which provide an alternative way of approaching the
star formation history. Finally, spectroscopic redshifts may be used 
to refine photometric redshift estimates and to test what
fraction of photometric redshifts are in error
in a given field. Some classes of sources, such as broad-line AGNs
or sources that have very strong emission lines, frequently have
bad photometric redshifts (e.g., see \markcite{barger03}Barger et al.\ 2003),
and spectroscopic measurements can be used to determine how common 
such sources are. The present paper, and a parallel paper by
\markcite{wirth04}Wirth et al.\ (2004), describe
two efforts to spectroscopically identify large, well-defined
samples in the ACS-GOODS region.

The region around the HDF-N was intensively observed spectroscopically
by a number of groups using the Low-Resolution Imaging Spectrograph
(LRIS; \markcite{oke95}Oke et al.\ 1995) on the Keck 10~m telescopes. 
This was a labor-intensive effort, since typically only $20-30$ sources 
could be observed in each mask, and $\sim 1$~hr observations 
were required.  However, spectra were obtained for 671 sources
(see the compilation of \markcite{cohen00}Cohen et al.\ 2000 for
full referencing). Analyses of various aspects of this sample are 
given in \markcite{cohen01}Cohen (2001) and \markcite{cohen02}Cohen (2002).

The recent implementation of the Deep Extragalactic Imaging Multi-Object
Spectrograph (DEIMOS; \markcite{faber03}Faber et al.\ 2003) on the 
Keck~II 10~m telescope has greatly increased the speed with which such 
observations may be carried out. DEIMOS can observe around 100 or
more sources per mask, and its higher resolution and
quantum efficiency make identifications more straightforward.
The long axis of the field-of-view of the instrument is
well matched to the 18 arcminute long axis of the ACS-GOODS field, and
most of the ACS-GOODS field can be covered in two DEIMOS pointings.
With DEIMOS it should be possible to identify nearly all of the sources
with optical AB magnitudes brighter than 24 in the red through
the entire ACS-GOODS area.

Recognizing this, two parallel efforts were begun
in Spring 2003 to map spectroscopically the galaxies in the ACS-GOODS
area with DEIMOS. A team led by Keck astronomers (hereafter, the Keck Team
Redshift Survey or KTRS) focused purely on an $R$ magnitude selected 
sample in the ACS-GOODS region, using observing and reduction techniques
developed for the DEEP-2 project 
(\markcite{davis03}Davis et al.\ 2003). This sample
is described in the paper by 
\markcite{wirth04}Wirth et al.\ (2004).
Our own sample was embedded within observations targeted
at high-redshift galaxies and X-ray and radio selected galaxies in
the wider regions, as described in
\markcite{hu04}Hu et al.\ (2004),
\markcite{barger03}Barger et al.\ (2003), and
\markcite{cowie04}Cowie et al.\ (2004). 
Free regions of the masks that lay within the ACS-GOODS
area were filled with magnitude selected samples
chosen to have $z'<23.5$, $R<24$, or $B<24.5$. Some
fainter sources were also included where no suitable
brighter targets could be obtained. Combining our DEIMOS data
with our LRIS data from earlier runs, we now have spectroscopic
identifications for 787 sources in the field with
$z'<24$, $R<24.5$, or $B<25$.
Combining this sample with data from the literature
(\markcite{cohen00}Cohen et al.\ 2000, hereafter C00; 
\markcite{dawson01}Dawson et al.\ 2001, hereafter D01; 
\markcite{steidel03}Steidel et al.\ 2003, hereafter S03) 
increases the identified sample to 1180.

In the present paper we describe our spectroscopic sample
and compare it with other spectroscopic and photometric redshift 
samples, focusing on our sample's completeness.
We then analyze our $z'<23.5$ selected sample 
in detail to look at how the colors and the star formation and AGN 
signatures of the galaxies depend on their density environment 
over the $z=0-1.5$ redshift range. 
We take $H_0=65~h_{65}$~km~s$^{-1}$~Mpc$^{-1}$ and use an
$\Omega_M={1\over 3}$, $\Omega_\Lambda={2\over 3}$ cosmology.

\section{Observations}
\label{secobs}

Since the ACS-GOODS observations were in progress at
the time of our mask designs, the astrometric
and photometric properties of the sample were taken
from the ground-based catalog of 
\markcite{capak03}Capak et al.\ (2003).
In the present paper we use the astrometric positions
from the ACS-GOODS catalog, which were kindly provided by
Mauro Giavalisco, but we continue to use the photometric
magnitudes from Capak et al.\ We chose the DEIMOS sample
from galaxies in the area with $z'<23.5$, 
supplemented by galaxies with $R<24$ or with $B<24.5$. 
Where no primary target could be fitted into a slit mask, 
we included sources up to half a magnitude fainter. 
As noted above, many of the sources already had redshifts in 
the literature or from our own previous LRIS spectroscopy. 
In planning our DEIMOS observations, we prioritized unobserved 
sources over those with existing redshifts. However, where 
no unobserved target was available, we included sources with
known redshifts, since one of our primary goals was to obtain 
as uniform a set of spectra for the sources as possible.

The new spectroscopic observations were obtained with DEIMOS
on Keck~II the nights of UT 2003 January 29--30, March 27, April
25--27, and May 27. The observations were made with the 600 lines 
per mm grating, giving a resolution
of $3.5$~\AA\ and a wavelength coverage of $5300$~\AA. The spectra
were centered at an average wavelength of $7200$~\AA, though the
exact wavelength range for each spectrum depends on the slit position.
Each $\sim 1$~hr exposure was broken
into three subsets, with the sources stepped along the slit by
$1.5''$ in each direction. The spectra were reduced in the same
way as previous LRIS spectra (see \markcite{cowie96}Cowie et al.\ 1996)
by forming a primary sky subtraction
from the median of the dithered images, registering and combining
the images with a cosmic ray rejection filter, removing geometric
distortion, and then applying a profile weighted extraction to
obtain the spectrum. The wavelength calibration was made using
a polynomial fit to the night sky spectrum.

Each spectrum was individually inspected, and a redshift was
measured only where a robust
identification was possible. The high resolution of the DEIMOS
observations greatly simplifies the identification of sources
dominated by single strong emission lines since
it resolves the doublet structure of the [OII]~3727~\AA\ line.
As discussed below, most of the magnitude-limited
sample sources could be identified. The incompleteness
is largest at the faint end, but in a small number
of cases (about $5-10$\%, depending on the mask) the spectra
were problematic in some way (e.g., the sources were at the edge 
of a CCD chip or in the heavily distorted and vignetted regions
at the edge of the mask); these are present at all magnitudes.
Sources that were observed but could not be identified (independent
of the cause) are retained in our main table (Table~1).

In Table~1 we list the 905 sources that we observed with either DEIMOS
or LRIS and have spectra for, from any part of the ACS-GOODS area, and 
satisfying the magnitude constraints $z'<24$, $R<24.5$, or $B<25$.
Table~1 is ordered by $R$ magnitude and gives the right ascension 
and declination in decimal 
degrees (from the ACS-GOODS catalog), the $z'$, $R$, and $B$ magnitudes 
(from the ground-based data of \markcite{capak03}Capak et al.\ 2003), 
and the measured redshift. The magnitudes are $3''$ diameter aperture 
magnitudes corrected to total magnitudes. Where a source is saturated 
or otherwise problematic, the magnitudes are omitted. In the 
redshift column, sources which were observed but 
not identified are shown as ``obs''. Of the sources in the table, 
787 are identified, comprising 704 galaxies and 83 stars.
We refer to the sample in Table~1 as the ``Hawaii'' sample.

In Table~2 we combined the identified sources in the 
``Hawaii'' sample with redshifts from the literature to form a 
more complete table. Including 
the data from C00, D01, and S03, we have redshifts or stellar
identifications for 1180 sources. 
Table~2 gives the right ascension 
and declination in decimal degrees from the ACS-GOODS catalog,
the $HK'$, $z'$, $I$, $R$, $V$, $B$, and $U$ magnitudes from
\markcite{capak03}Capak et al.\ (2003), the redshift and its source 
(H=Hawaii, C00, D01, or S03), the photometric redshift from 
\markcite{capak04}Capak et al.\ (2004), and the odds assigned 
to it (see \S~\ref{secphot}). Sources 
in the sample that are compact in the ACS-GOODS images and are
classified as stars from the empirical star-galaxy separation 
criterion $(V-I)-2.0(I-HK')=-1.1$ given in 
\markcite{barger99}Barger et al.\ (1999) are assigned as stars 
in the photometric redshift list. We refer to the
sample in Table~2 as the ``total'' sample.

There are a very small number of discrepant sources, where 
different groups have assigned different redshifts. In only one 
instance does a Hawaii DEIMOS redshift differ from a Hawaii LRIS 
redshift. In this case, the DEIMOS redshift also turns out to be 
inconsistent with other groups' measurements (see Table~3),
so we adopt the LRIS redshift in our ``Hawaii'' sample.

\section{Tests}
\label{sectest}

\subsection{Comparison with other catalogs}
\label{secother}

In order to compare the ``Hawaii'' sample with the KTRS sample of
\markcite{wirth03}Wirth et al.\ (2003) and the literature samples 
(C00, D01, and S03), we first combined the KTRS and literature samples
to form the ``other'' sample. 
In the small number of cases where there were discrepancies, we somewhat 
arbitrarily resolved these in favor of KTRS, then C00, then D01,
and then S03. There are 552 overlapping identifications between
the ``Hawaii'' sample and the ``other'' sample. Of these, 512
are galaxies and 40 are stars. All of the star identifications
are consistent in the two samples. In Figure~\ref{fig1} we compare the 
``Hawaii'' galaxy redshifts with the ``other'' galaxy redshifts and find
that nearly all of these are also consistent. Considering a redshift to be 
discrepant if it deviates by more than 0.01, we find that only seven of 
the overlap galaxies are discrepant, in addition to the one internally 
inconsistent galaxy discussed in \S~\ref{secobs}; the latter source is not 
shown in Fig.~\ref{fig1} because we adopted our LRIS redshift instead of 
our DEIMOS redshift. These eight galaxies give an upper bound of 
1.4\% of sources where the identifications might be problematic in the 
``Hawaii'' sample; they are summarized and briefly described in Table~3.

%
%
\begin{inlinefigure}
\centerline{\psfig{figure=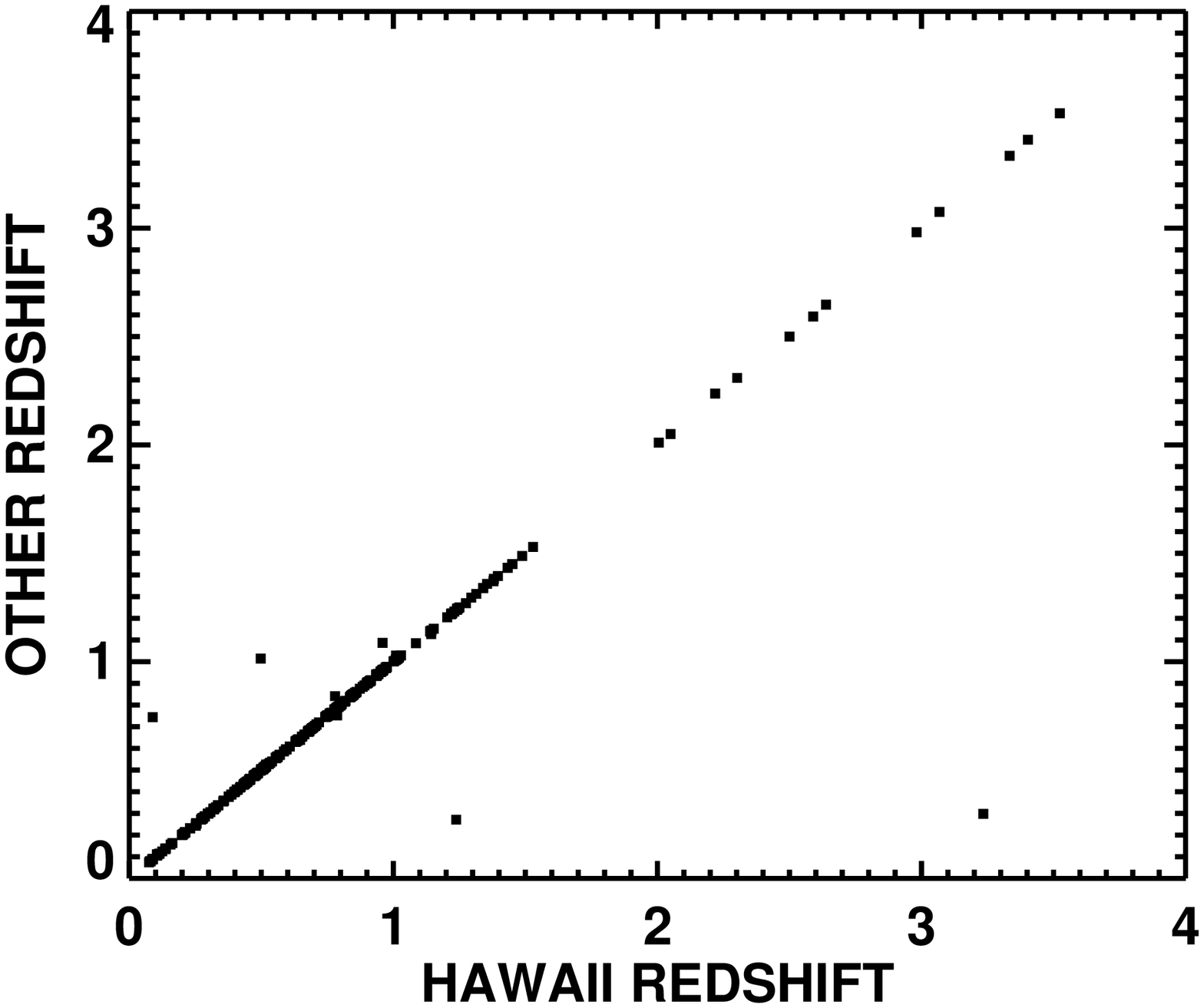,width=3.2in,angle=90}}
\caption{
``Hawaii'' spectroscopic redshift versus ``other'' spectroscopic 
redshift (from the KTRS and literature samples). There are
512 overlapping galaxy identifications between the ``Hawaii'' 
sample and the ``other'' sample, only seven of which have discrepant 
redshifts.
\label{fig1}
}
\addtolength{\baselineskip}{10pt}
\end{inlinefigure}

\subsection{Photometric redshift comparison}
\label{secphot}

We next compared the ``total'' sample of 
redshifts (from the Hawaii observations and the literature)
with the photometric redshift catalog of 
\markcite{capak04}Capak et al.\ (2004).
The photometric redshift estimates are based on ultradeep imaging 
in seven colors from $U$ to $HK'$ 
(\markcite{capak03}Capak et al.\ 2003) and were computed using 
the Bayesian code of 
\markcite{benitez00}Ben\'{\i}tez (2000). This code
assigns odds that the redshift lies close to the estimated
value. We only use those cases where the odds
lie above 90\%. The imaging data also saturate at bright
magnitudes, so we restrict the comparison to sources
with $R>20$. This gives a sample of 1078 sources with
both spectroscopic and photometric redshifts, which
we compare in Figure~\ref{fig2}. 
Fourteen of the sources are
broad-line AGNs, where the photometric estimates may be poorer
since the templates do not properly model such sources.
We have distinguished these sources in Figure~\ref{fig2}
with larger solid squares. 97\% of the photometric redshifts agree 
to within a multiplicative factor of 1.3 with the spectrosopic
redshifts. This is consistent with the expectation from the odds 
assigned to the photometric redshifts. The agreement is independent
of magnitude over the $R=20-24$ range.

%
%
\begin{inlinefigure}
\centerline{\psfig{figure=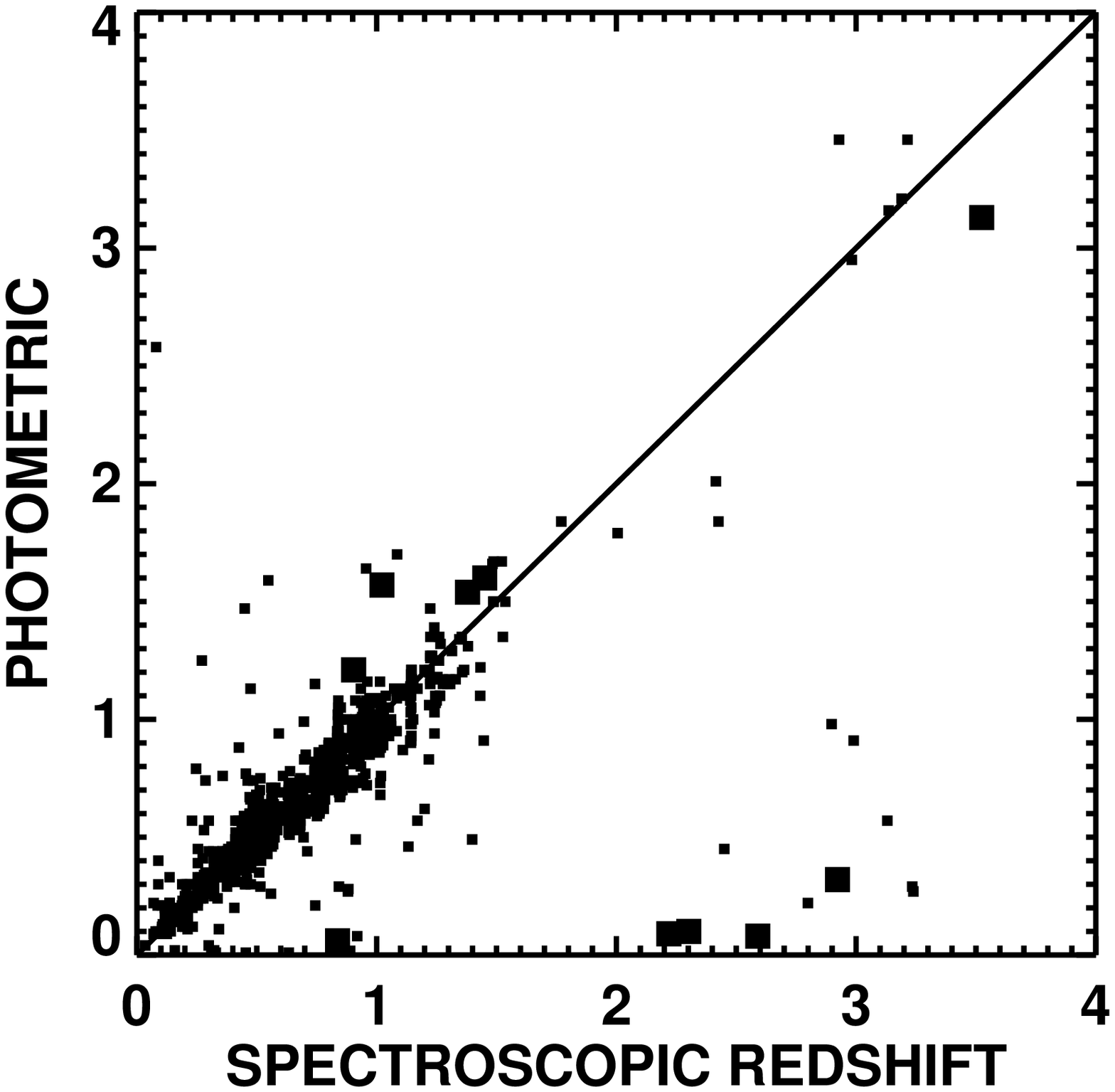,width=3.2in,angle=90}}
\caption{
Photometric redshift versus spectroscopic redshift
for the 1078 sources from the ``total'' sample
with $20<R<24$ having both redshift 
measurements. The 14 broad-line AGNs are denoted
by large solid squares. Only 38 of the photometric
redshifts differ by more than a multiplicative
factor of 1.3 from the spectroscopic redshifts;
eight of these are broad-line AGNs.
\label{fig2}
}
\addtolength{\baselineskip}{10pt}
\end{inlinefigure}

We may also use the photometric redshifts to test whether
there are redshift intervals where the spectroscopic identifications
are substantially incomplete. In Figure~\ref{fig3} we compare for
three magnitude bins the histogram ({\it dotted line})
of the spectroscopic redshifts (from the ``total'' sample) with the 
histogram ({\it dashed line}) of the photometric redshifts.
In the $R=20-22$ and $R=22-23$ bins, the shapes of the two histograms
are similar, and an appropriate renormalization ({\it solid line}) 
produces close agreement between the two distributions.
However, in the $R=23-24$ interval, a higher fraction
of sources are expected to lie at $z>1.5$ on the basis of 
the photometric redshifts than are spectroscopically
observed at such redshifts.
This is the well-known spectroscopic ``desert'', where the 
[OII]~3727~\AA\ line has moved out of the optical
window, while the ultraviolet signatures have not yet entered.

%
%
\begin{inlinefigure}
\centerline{\psfig{figure=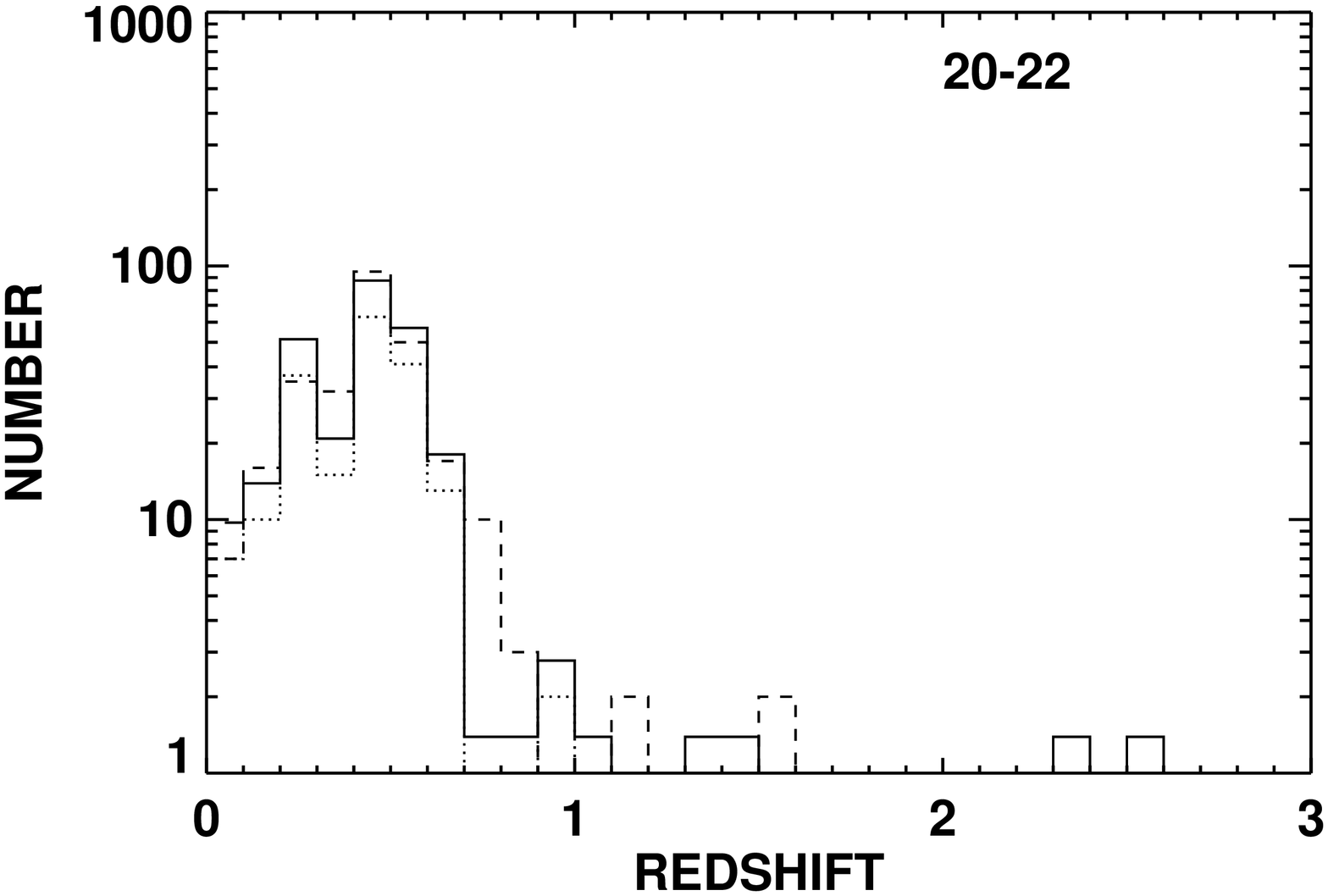,width=3.2in,angle=90}}
\centerline{\psfig{figure=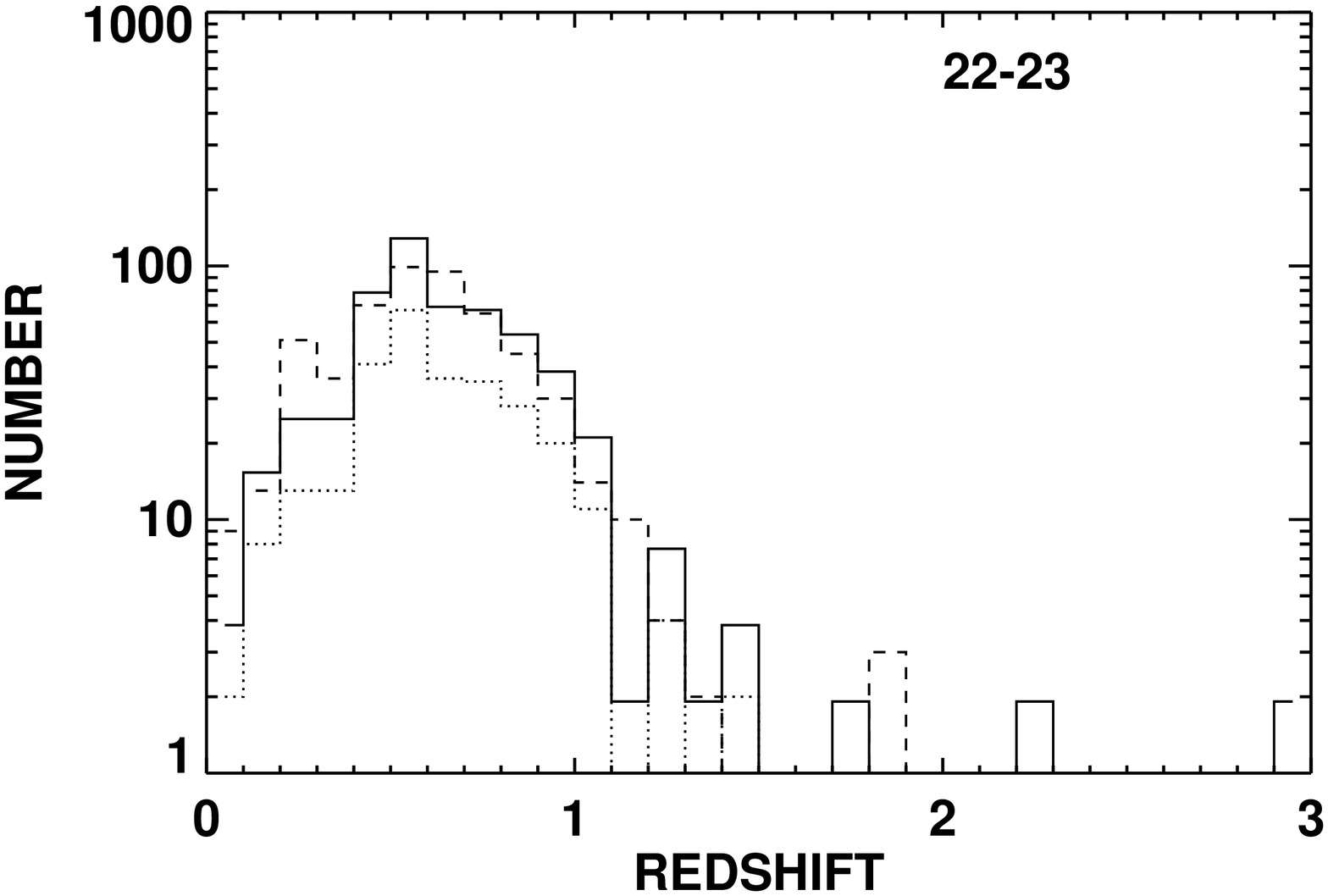,width=3.2in,angle=90}}
\centerline{\psfig{figure=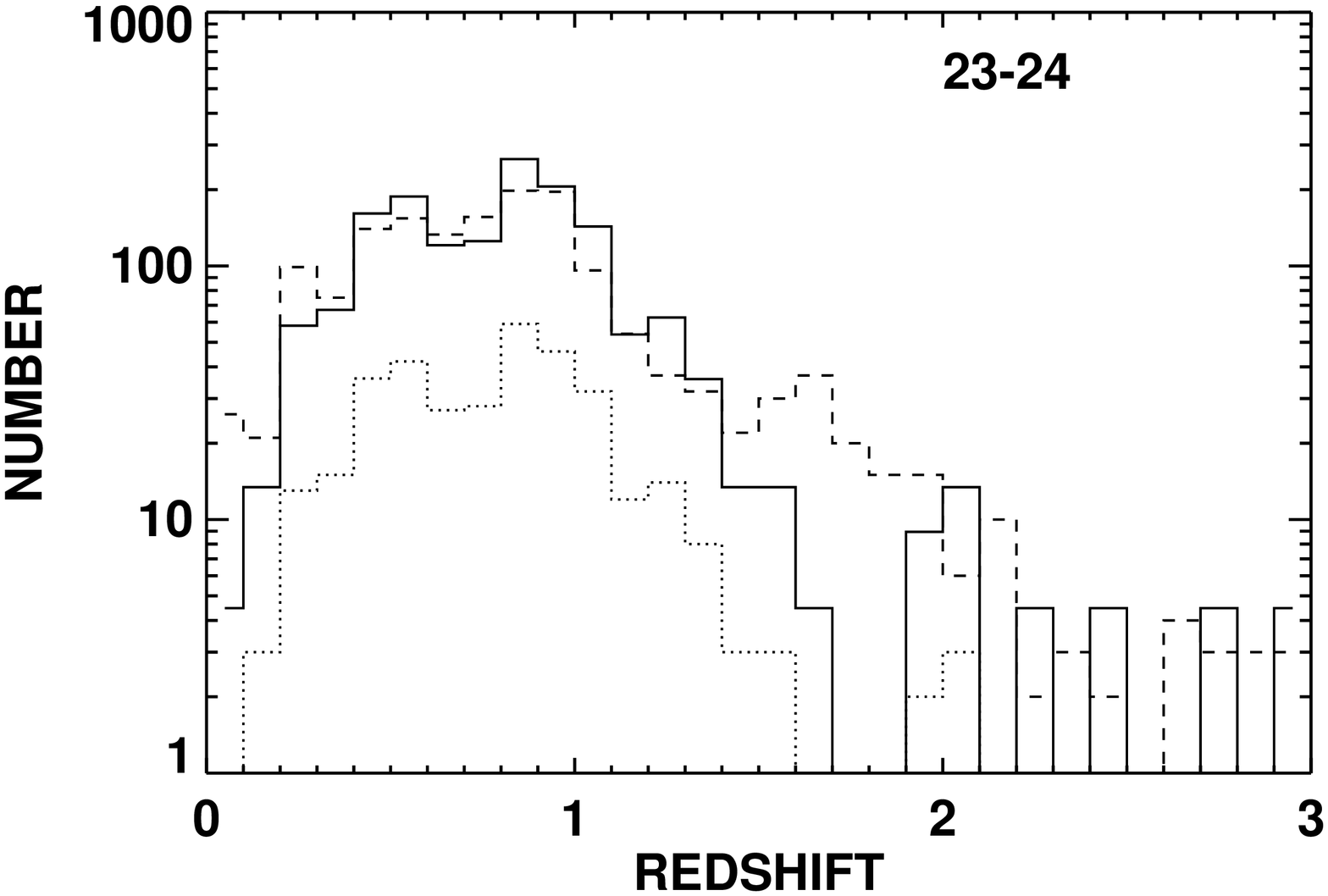,width=3.2in,angle=90}}
\caption{Observed spectroscopic
(from the ``total'' sample; {\it dotted})
and photometric ({\it dashed}) redshift distributions
for three magnitude bins: (a) $R=20-22$, (b) $R=22-23$, 
and (c) $R=23-24$. Solid line
shows the spectroscopic distribution renormalized to
match the total number of sources. Only region
of significant incompleteness is the spectroscopic desert 
at $z=1.5-2$ in the faintest magnitude ($R=23-24$) interval. 
\label{fig3}
}
\addtolength{\baselineskip}{10pt}
\end{inlinefigure}

%
%
\begin{inlinefigure}
\centerline{\psfig{figure=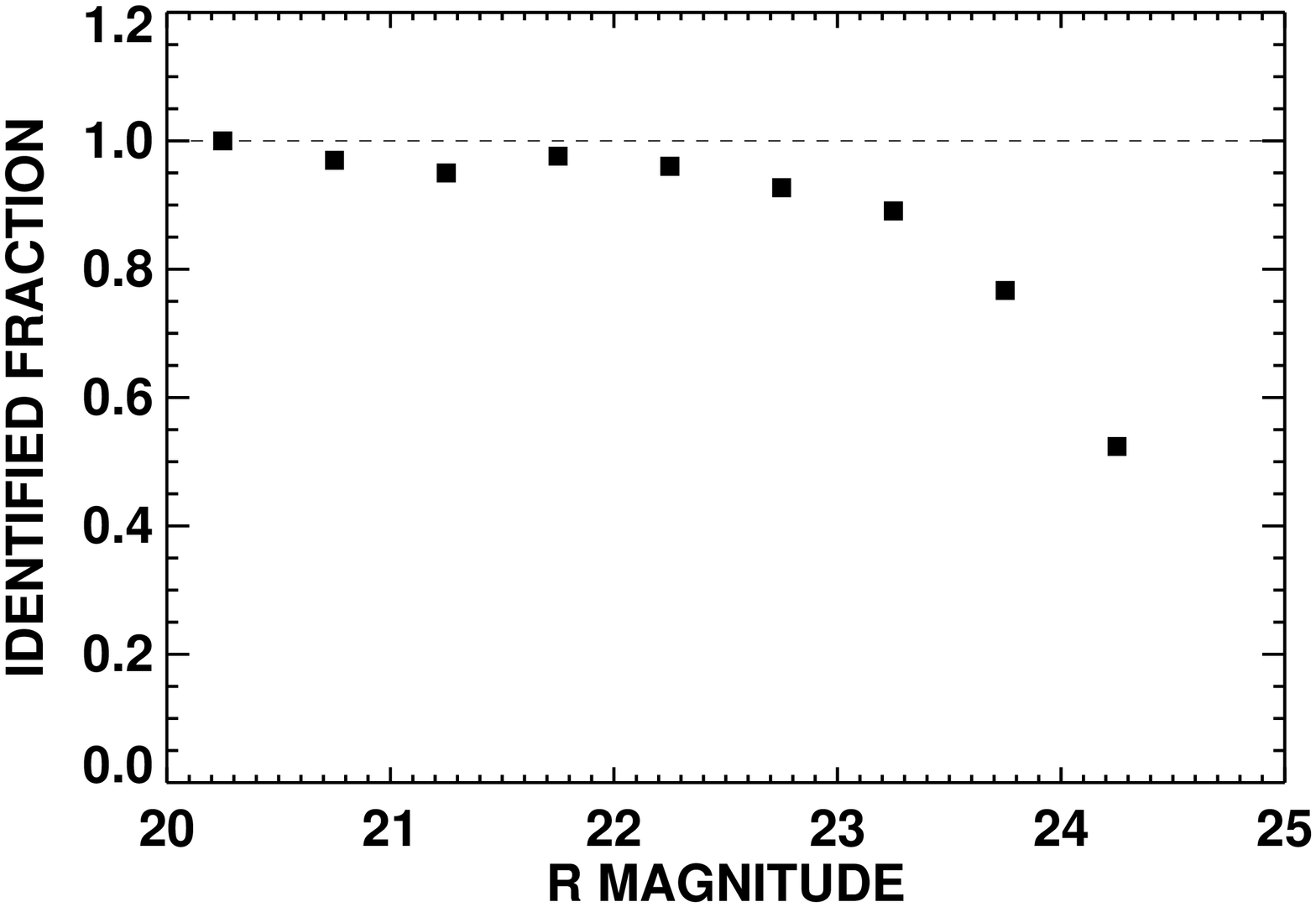,width=3.2in,angle=90}}
\centerline{\psfig{figure=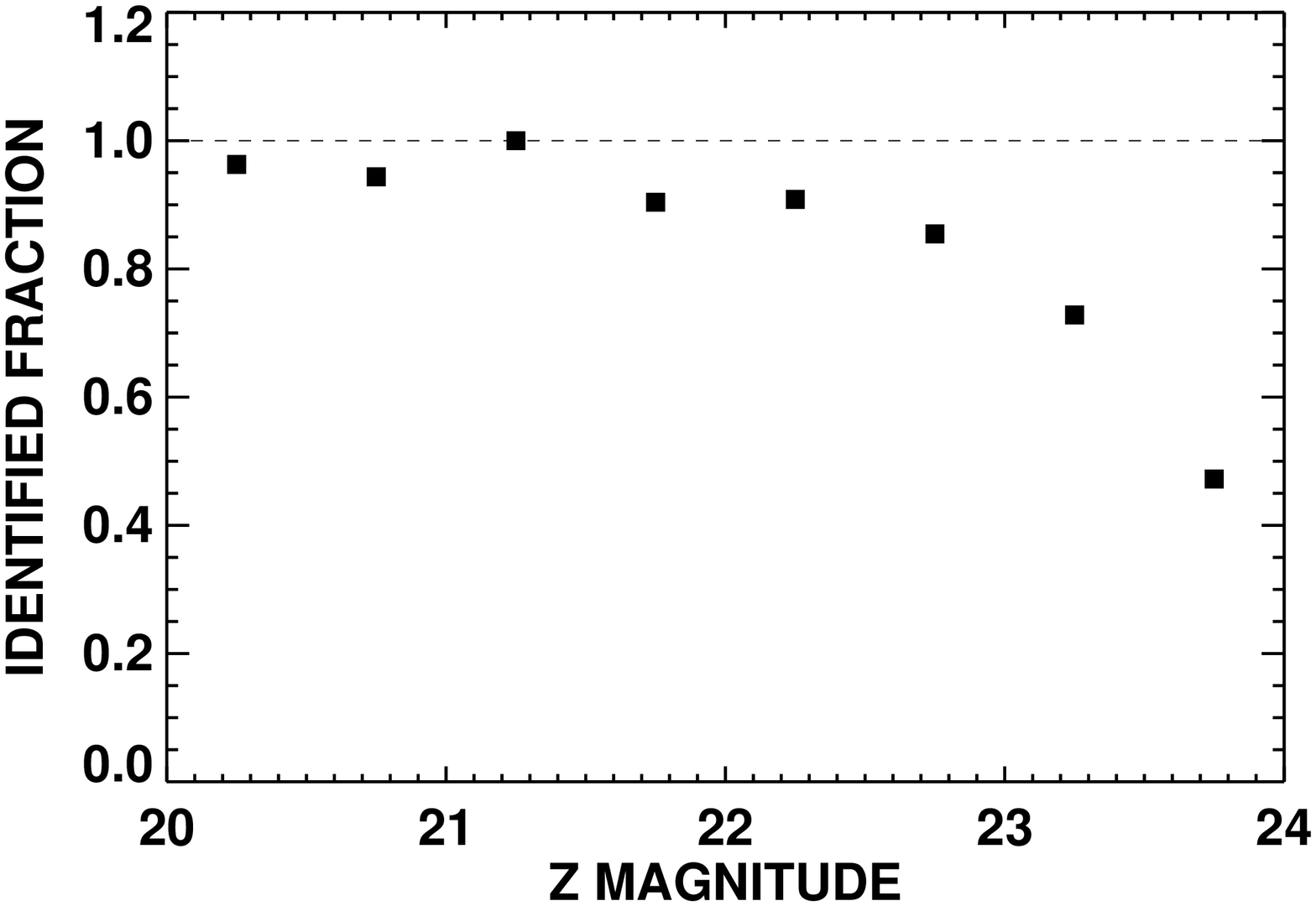,width=3.2in,angle=90}}
\caption{
Fractional completeness in 0.5~mag intervals
of the ``Hawaii'' DEIMOS spectroscopic identifications using
(a) the $R$ selected sample and (b) the $z'$ selected sample.
\label{fig4}
}
\addtolength{\baselineskip}{10pt}
\end{inlinefigure}

\subsection{Completeness}
\label{seccomp}

We next tested the fractional completeness of the ``Hawaii'' 
DEIMOS spectroscopic identifications versus magnitude.
The results, which should be typical of what can be achieved
with 1~hour exposures on this instrument,
are given in Figure~\ref{fig4}.
Below $R=23$, nearly all of the observed sources
are identified, except in a very small fraction of cases
where there were problems with the spectra.
The completeness begins to drop off beyond $R=23$, and by $R=24$,
it has fallen to about 60\%. Some part of this may be a consequence
of there being a higher fraction of $z>1.5$ sources at these
fainter magnitudes, since such sources in the spectroscopic
desert are hard to identify. The photometric redshift
estimates shown in Figure~\ref{fig3} predict that 13\% of the
sources in the $R=23.75-24.25$ interval should lie between
$z=1.5$ and $z=2$, while only 5 sources are spectroscopically
identified in this redshift interval. This suggests that about
a third of the incompleteness is caused by redshift
evolution moving the sources into difficult redshift ranges
and that the remaining incompleteness is primarily caused by the 
lower S/N in the spectra of the fainter sources.

%
%
\begin{inlinefigure}
\centerline{\psfig{figure=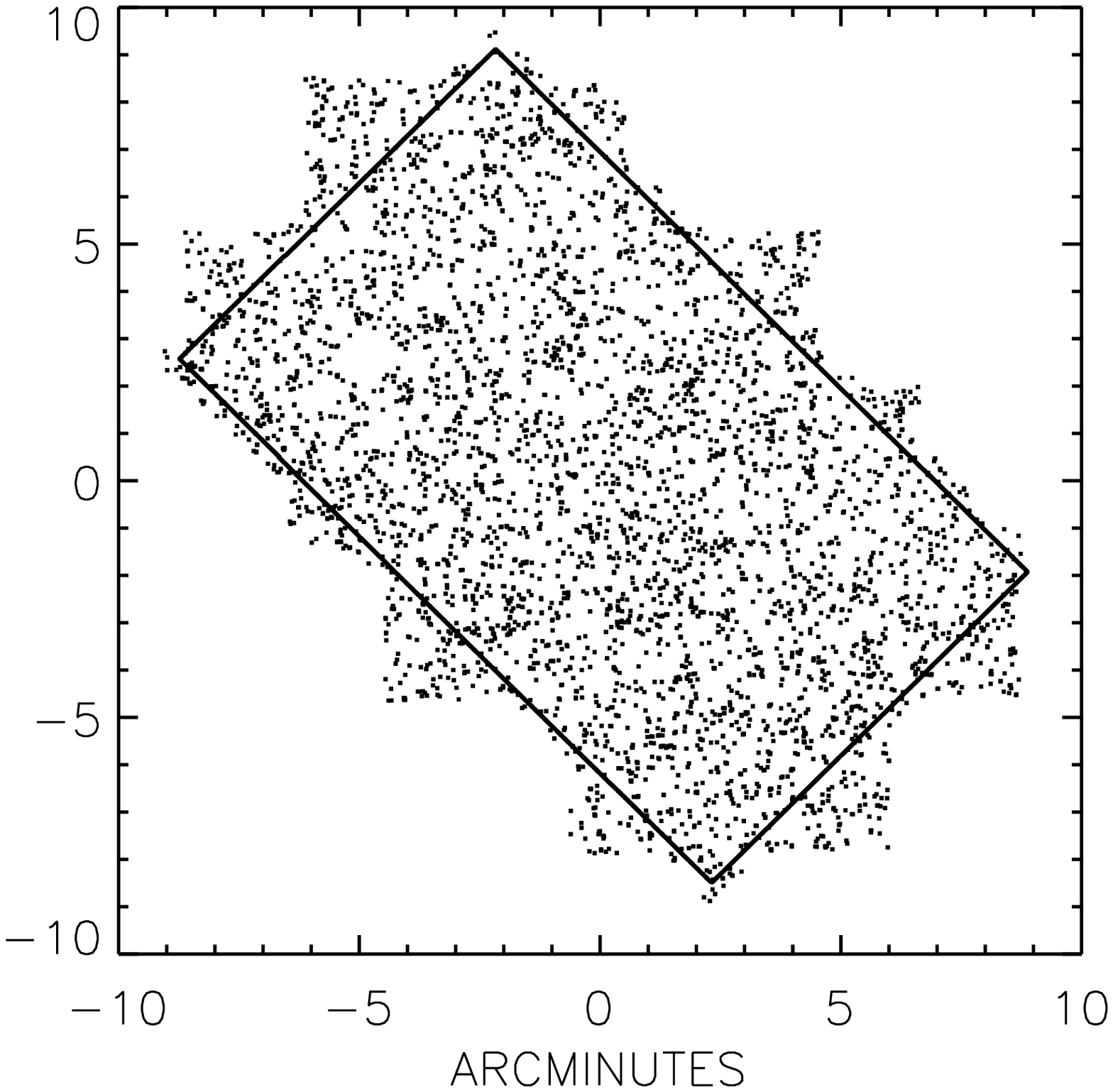,width=3.2in,angle=90}}
\caption{
Magnitude selected sources in the ACS-GOODS region.
To simplify the complex geometry and eliminate sources that are 
poorly covered in the ACS tiling, we restrict to 
sources lying within the rectangular area shown. Rectangle is
centered at 189.227018 and 62.238091, with a long axis of
15.63~arcmin and a short axis of 9.28~arcmin.
The long axis is oriented at 45 degrees to the N-S direction
and runs from NE to SW.
\label{fig5}
}
\addtolength{\baselineskip}{10pt}
\end{inlinefigure}

\subsection{Fraction Observed}

The ACS-GOODS region has a complex morphology due to the tiling 
process (see Fig.~\ref{fig5}). To eliminate the more poorly 
covered regions, from now on we restrict our
analysis to the rectangular 145~arcmin$^2$ area shown in
Figure~\ref{fig5}; hereafter we refer to this as the ``restricted''
field. The coordinates and size of this area are 
given in the Figure~\ref{fig5} caption. The area contains 3460 
sources that satisfy the primary magnitude selection criteria 
($z'<23.5$, $R<24$, or $B<24.5$). 

Figure~\ref{fig6} shows the fractional identifications versus
$R$ and $z'$ magnitude for the ``total'' sample in the restricted
field. Figure~\ref{fig7} shows the 
same versus position along the long axis of the rectangle.
More than a third of the sources have been identified to $R=24$
and $z'=23.5$, with the fractions being higher at the
brighter magnitudes. The ``total'' sample is rather 
strongly concentrated around the HDF-N proper (see Fig.~\ref{fig7}). 

In order to show the current completeness of all spectrscopic
data on the field we have
combined the present data with the
KTRS sample of Wirth et al. (2003). In the restricted field this
raises the number 
of identified sources to slightly more than 1800, or more than 
half the sources to $R=24$ (see Fig.~\ref{fig6}). The inclusion
of the KTRS data also substantially smooths the spatial
distribution (see Fig.~\ref{fig7}).

%
%
\begin{inlinefigure}
\centerline{\psfig{figure=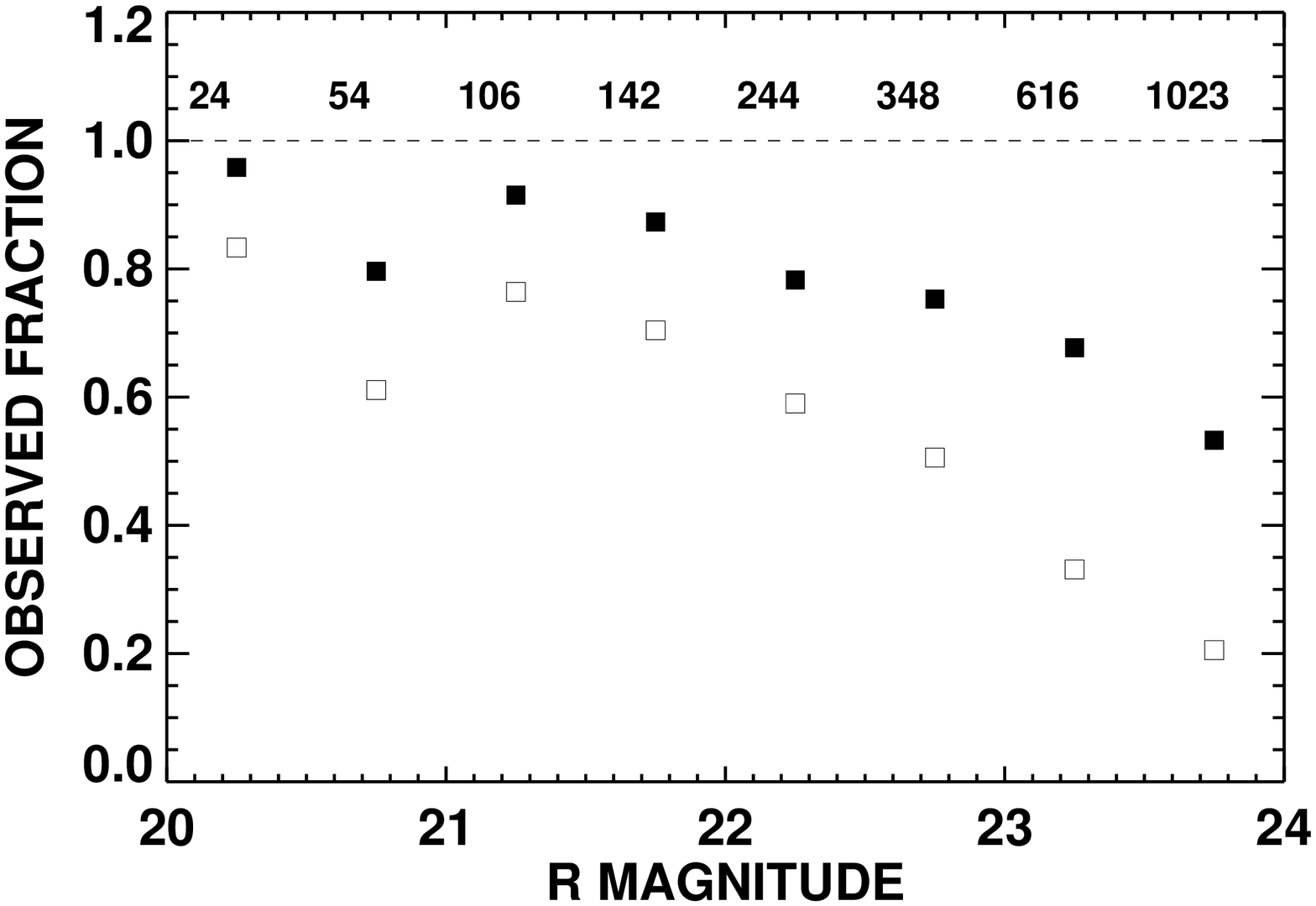,width=3.2in,angle=90}}
\caption{
Fractional completeness ({\it open squares}) of the $R<24$ ``total'' 
sample in the restricted field (as compared to all of the sources 
in the restricted field, whether observed or not) vs. $R$ magnitude.
Total number of sources in each magnitude bin is shown along the top.
Solid squares show the fractional completeness when combined with 
the KTRS sample in the restricted field. 
\label{fig6}
}
\addtolength{\baselineskip}{10pt}
\end{inlinefigure}

%
%
\begin{inlinefigure}
\centerline{\psfig{figure=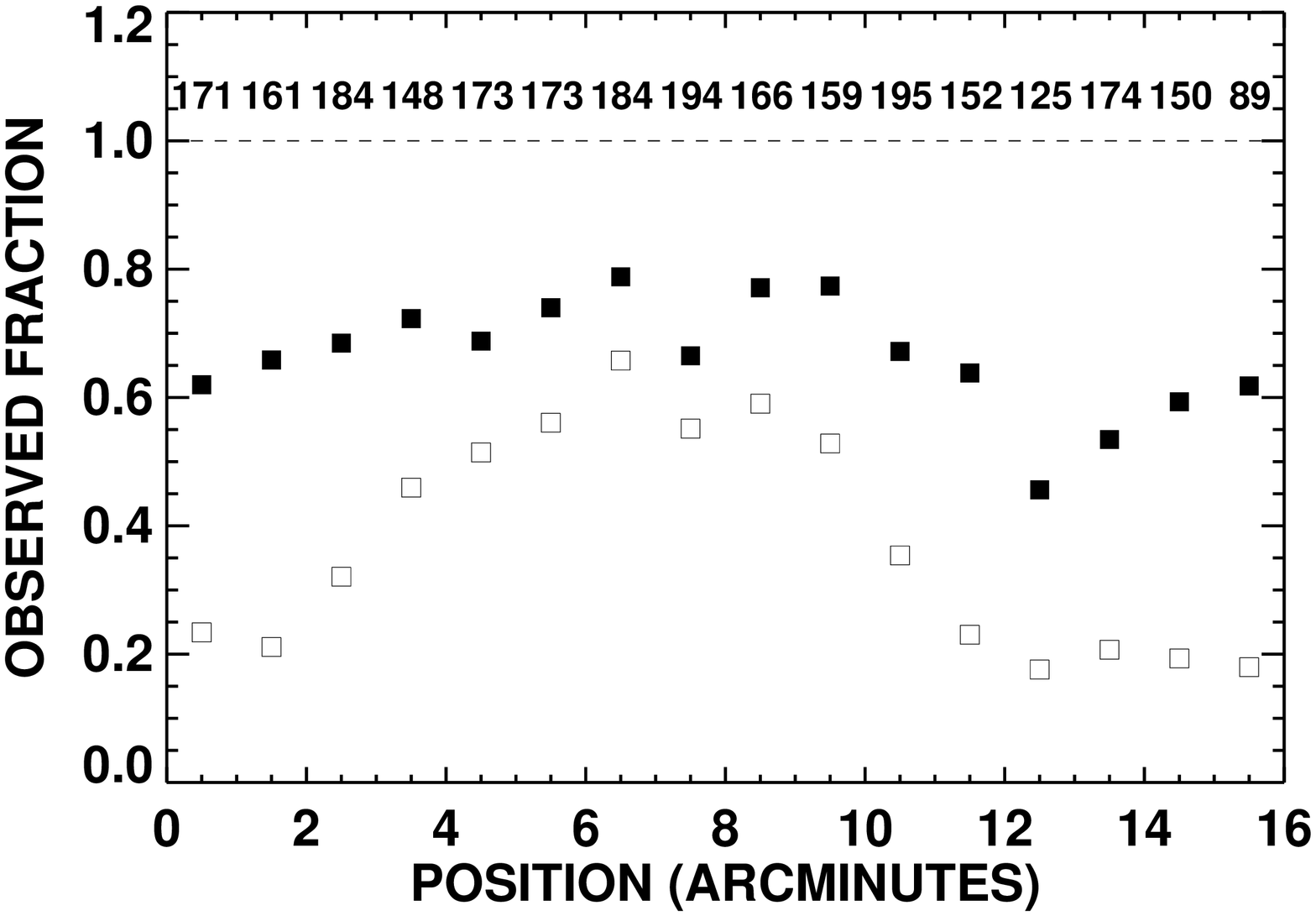,width=3.2in,angle=90}}
\caption{
Fractional completeness ({\it open squares}) of the $R<24$
``total'' sample in the restricted field
(as compared to all of the sources in the restricted 
field, whether observed or not) vs. position along 
the long axis of the ACS-GOODS field.  Total number of sources 
at each position is shown along the top. Solid squares show the 
fractional completeness when combined with the KTRS sample in the 
restricted field. 
\label{fig7}
}
\addtolength{\baselineskip}{10pt}
\end{inlinefigure}

\section{Discussion}
\label{secdisc}

\subsection{Redshift distribution}
\label{secredd}

Figure~\ref{fig8} shows the redshift-magnitude diagrams for the 
$R<24$ and $z'<23.5$ selected sources in the ``total'' sample 
in the restricted field. At magnitudes $R<23$
or $z'<22.5$, nearly all of the sources at redshifts higher
than $z=1.5$ are broad-line AGNs. At fainter magnitudes,
the more normal galaxy population stretches beyond
$z=1.5$ into the redshift desert (see \S~\ref{seccomp}),
and the truncation of the redshift distribution at $z=1.5$
is clearly seen. A small number of galaxies have measured
redshifts above $z=2$. A substantial fraction of the $z>2$
sources have X-ray counterparts ({\it large open squares}) and may
have substantial AGN contributions to their light, but many
do not (see \markcite{barger03}Barger et al.\ 2003).

%
%
\begin{inlinefigure}
\centerline{\psfig{figure=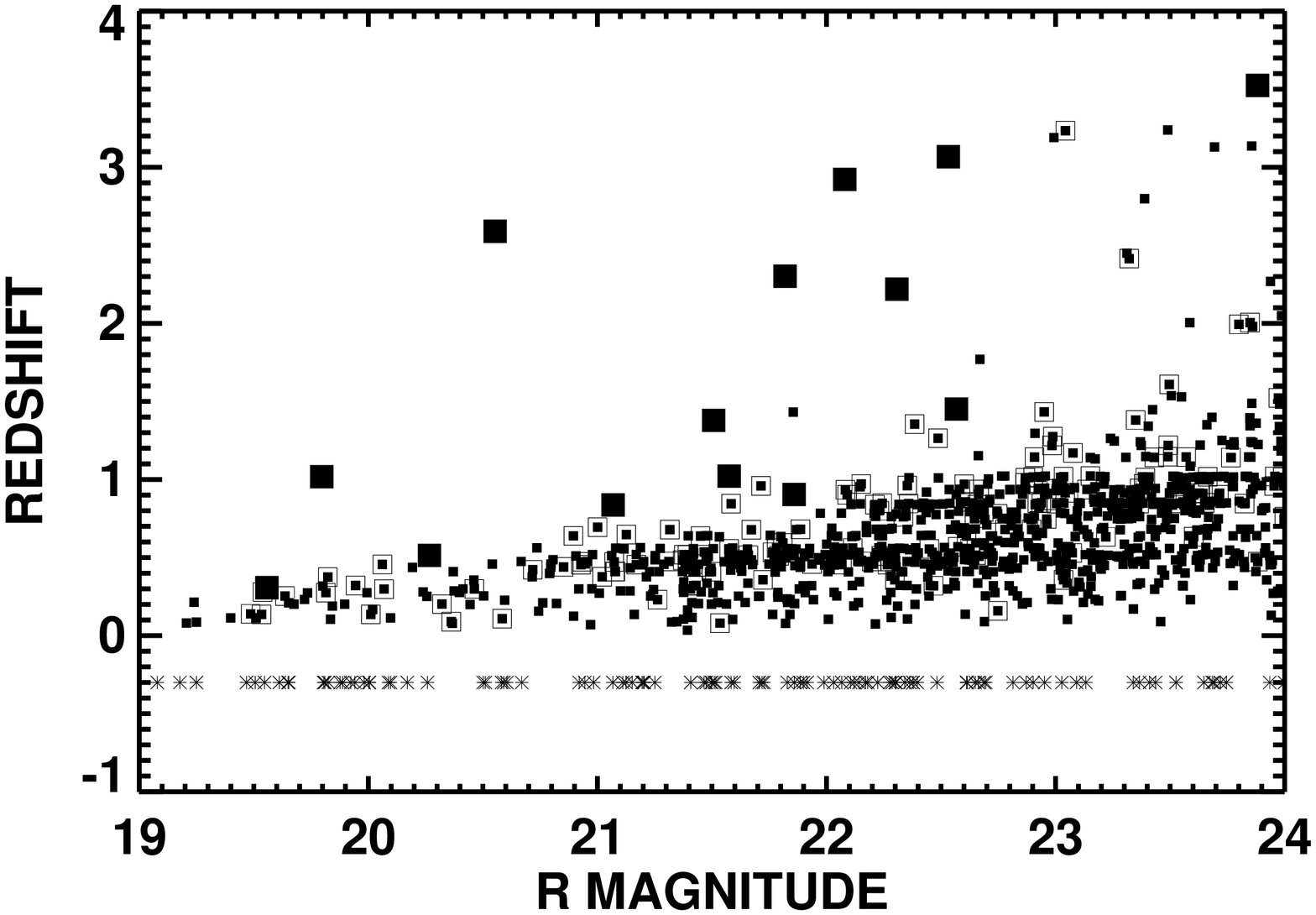,width=3.2in,angle=90}}
\centerline{\psfig{figure=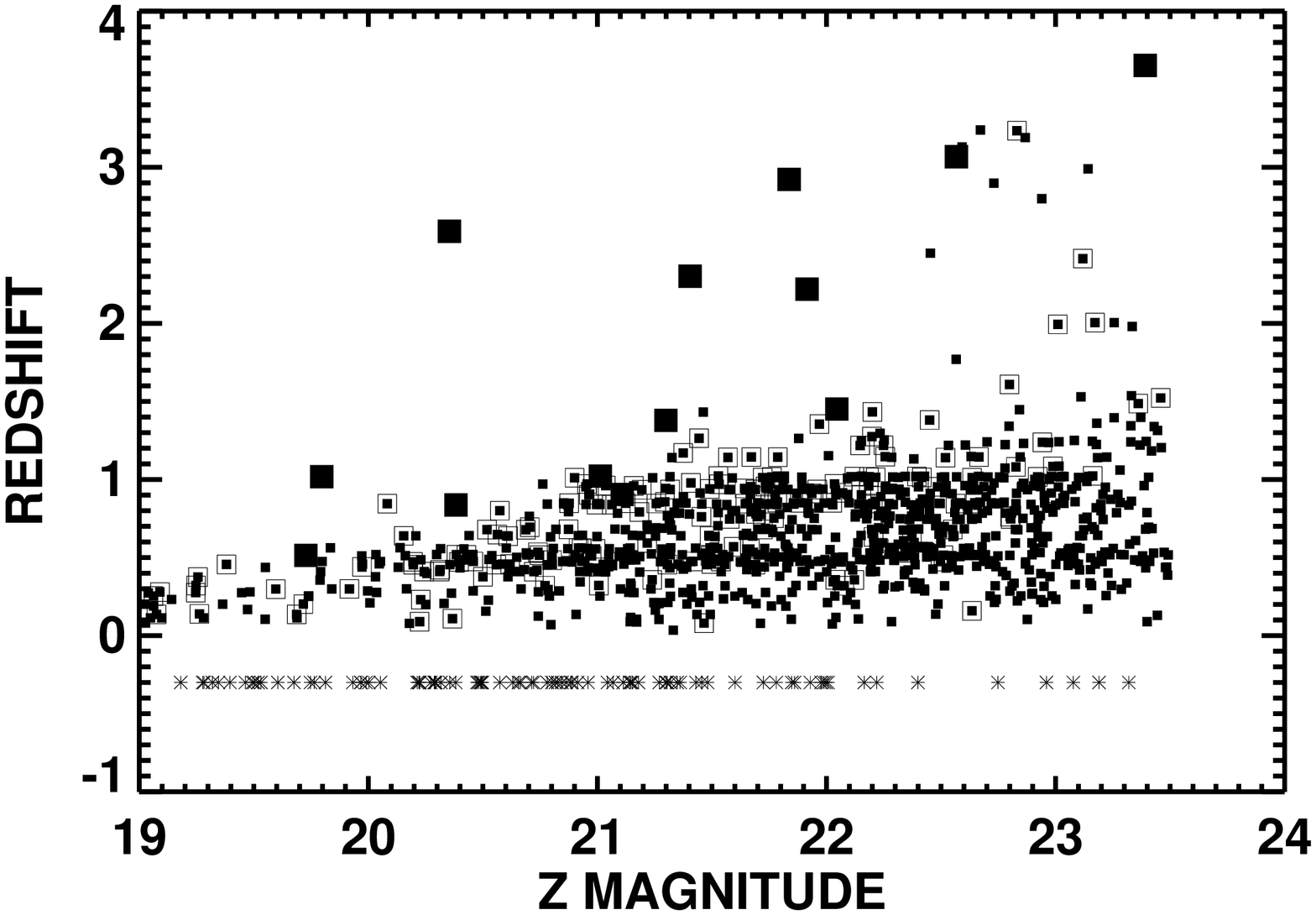,width=3.2in,angle=90}}
\caption{
Spectroscopic redshift versus magnitude for the (a) $R<24$
and (b) $z'<23.5$ selected sources in the ``total'' sample 
in the restricted field. Broad-line sources are denoted by
large solid squares, and X-ray sources by
large open squares. Stars are denoted by asterisks.
\label{fig8}
}
\addtolength{\baselineskip}{10pt}
\end{inlinefigure}

%
%
\begin{inlinefigure}
\centerline{\psfig{figure=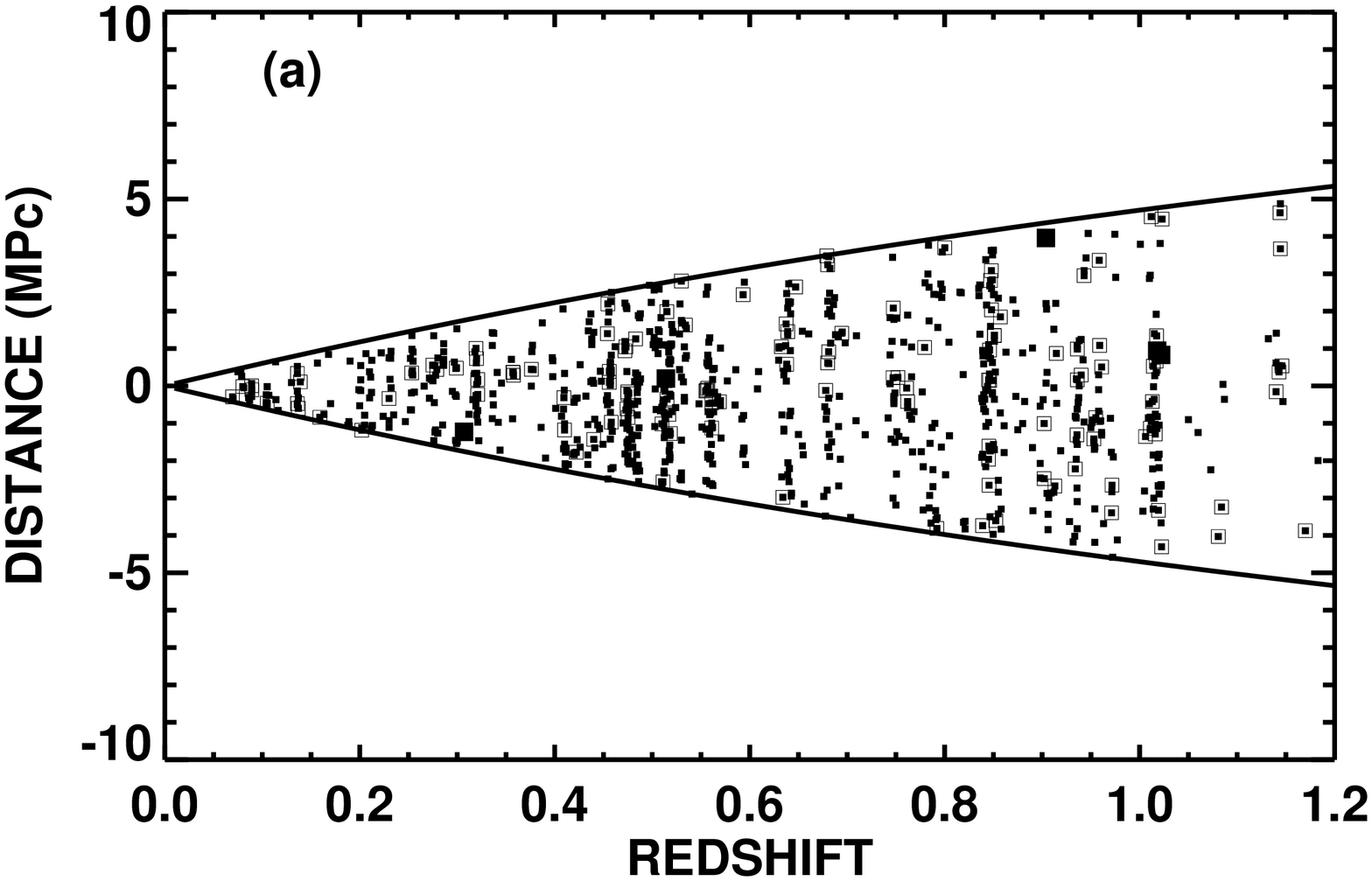,width=3.2in,angle=90}}
\centerline{\psfig{figure=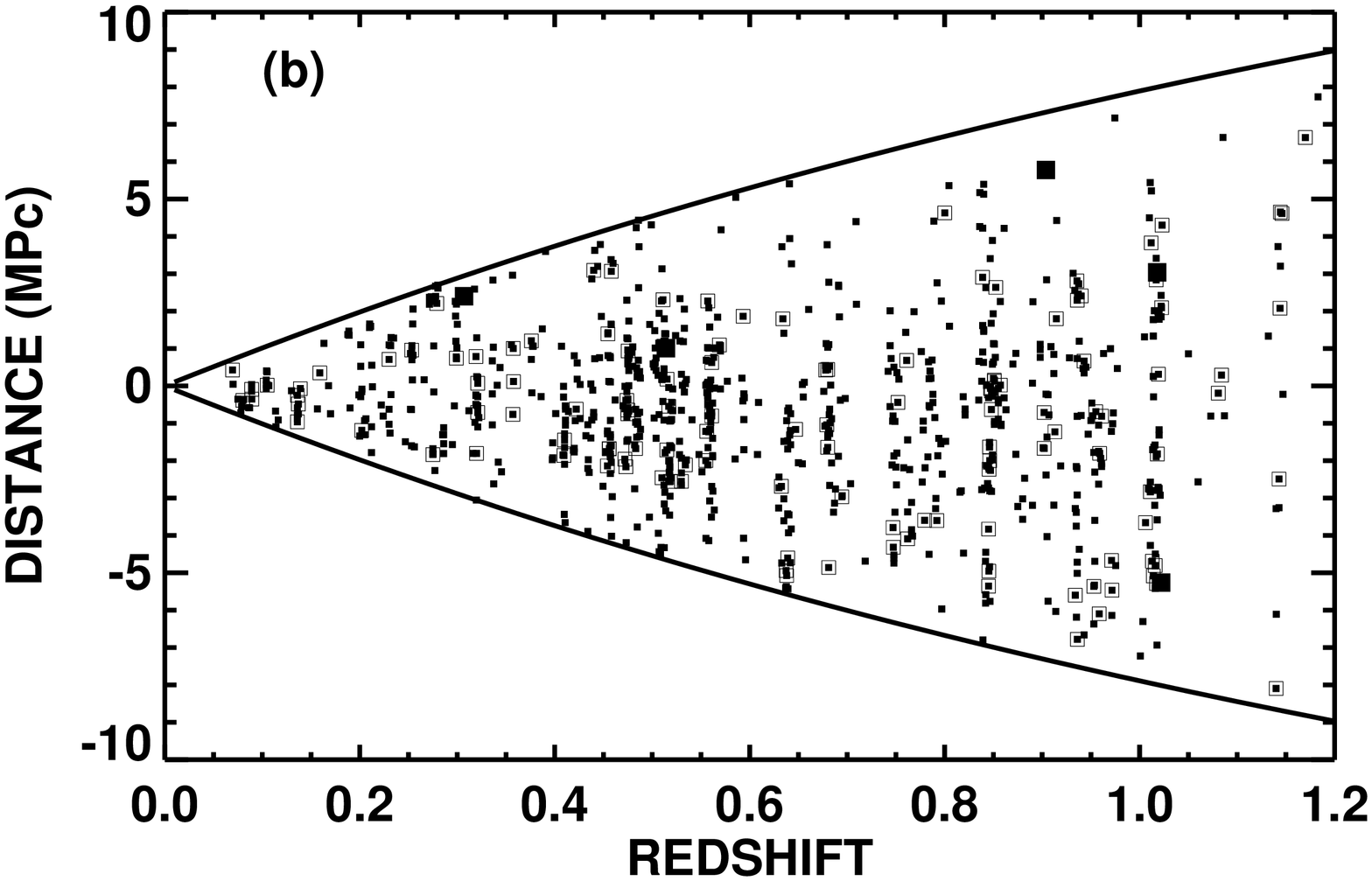,width=3.2in,angle=90}}
\caption{Pie diagrams for the spectroscopically identified
sources in the ``total'' sample in the restricted field.
(a) Slice parallel to the long axis of the ACS-GOODS
region. (b) Slice parallel to the short axis.
Angular distances are plotted versus redshift.
Broad-line sources are denoted by large solid squares, 
and X-ray sources by large open squares. All of
the broad-line AGNs are also X-ray sources.
\label{fig9}
}
\addtolength{\baselineskip}{10pt}
\end{inlinefigure}

The well-known ``velocity sheets'' (C00) that are a consequence 
of the large scale clustering can be seen in the redshift-magnitude 
diagrams. These are more clearly visible in the pie diagrams
of Figure~\ref{fig9}, which show the distributions (a) parallel 
and (b) perpendicular to the long axis of the ACS-GOODS rectangle. 
The most prominent of the sheets lie at $z=0.85$ and $z=1.01$. 
The sheets extend most of the way along the long axis (the drop-off 
at the ends is a consequence of the higher incompleteness
of the observations at these positions---see Figure~\ref{fig7}),
but the $z=1.01$ sheet is primarily found on the eastern
side of the rectangle. The X-ray source positions 
({\it large open squares}) and broad-line AGN positions 
({\it large solid squares}) are shown on the pie diagrams. 
As has been previously noted, these also show velocity sheet
structures (\markcite{barger02}Barger et al.\ 2002, 2003; 
\markcite{gilli03}Gilli et al.\ 2003). 
We shall return to the question of whether they are more 
concentrated into velocity sheets than the normal galaxies 
in \S~\ref{secx}.

\subsection{Galaxy colors versus redshift}
\label{seccol}

Figure~\ref{fig10} shows $R-z'$ color versus redshift for the 
$z'<23.5$ ``total'' sample in the restricted field. 
The upper bound of the distribution
matches extremely well to the expected color of an elliptical
galaxy. Apparently, very few galaxies are sufficiently reddened
to lie above this envelope. There are only six such sources,
three of which are galaxy pairs and radio sources 
({\it solid squares}), suggesting they may be dusty mergers. 
It is unclear why the other three sources lie at these colors.

%
%
\begin{inlinefigure}
\centerline{\psfig{figure=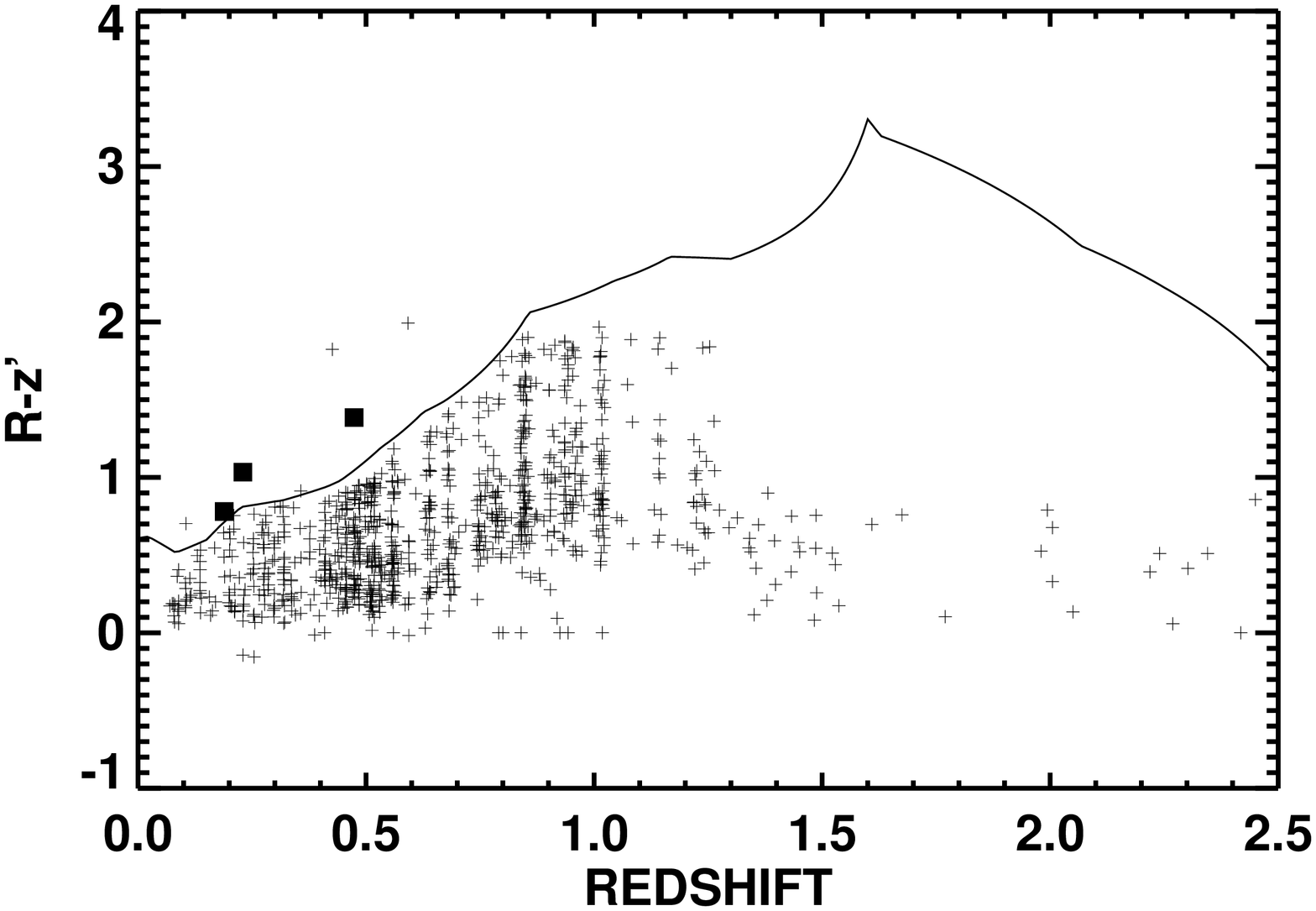,width=3.2in,angle=90}}
\caption{
$R-z'$ color versus redshift for the $z'=23.5$ ``total'' sample
in the restricted field. Only sources 
with spectroscopic redshifts are shown. Upper curve shows the color 
of the Coleman, Wu, \& Weedman (1980) elliptical galaxy 
versus redshift, which almost perfectly maps the upper envelope of 
the distribution over the $z=0-1$ redshift range. Only 6 sources lie
above the envelope, three of which are galaxy pairs and radio
sources ({\it solid squares}).
\label{fig10}
}
\addtolength{\baselineskip}{10pt}
\end{inlinefigure}

Because of the extensive color information available
(Table~2), we may analyze the galaxy colors in a more 
model-independent fashion by interpolating to form the rest-frame colors
at each redshift. In Figure~\ref{fig11} we show the rest-frame 
$2800-8000$~\AA\ color versus redshift. The upper bound matches 
the color of an elliptical galaxy and is nearly invariant from $z=0-1$.
Beyond this redshift, the $z'$ selection becomes biased
against elliptical galaxies as the 4000~\AA\ break enters
the $z'$ band. The bottom of the distribution corresponds
to irregulars, star formers, and broad-line AGNs (the latter
are denoted by large solid squares). 

The type selection versus absolute magnitude
is illustrated in Figure~\ref{fig12}, where we show the
rest-frame absolute (a) 8000~\AA \ and (b) 2800~\AA\ magnitudes
versus redshift. For Figure~\ref{fig12}a we use
$z'<23.5$, but we implicitly have a second selection
since we need to interpolate between the $HK'$ magnitudes
and the $z'$ magnitudes to determine the 8000~\AA\ magnitudes
at high redshifts. 
To formalize this, we also limit the sample to $HK'<22.5$. 
This $HK'$ selection is illustrated with a dashed curve computed 
for a flat-spectrum source. Above this curve, all sources should 
be included, if we assume that all sources are redder than
flat $f_{\nu}$. However, the $z'$ selection means that
we will omit redder sources at the higher redshifts
and fainter magnitudes. We illustrate
this for the case of an elliptical galaxy ({\it solid curve})
and a spiral galaxy ({\it dashed curve}). At $z=1.2$ we will omit
ellipticals fainter than about $-22$ in the rest-frame 8000~\AA\ band.
For Figure~\ref{fig12}b we use $R<24$.
Since this is a red-selected sample, we again illustrate
the selection threshold for a flat $f_{\nu}$ spectrum.
All sources above this line should be included.

%
%
\begin{inlinefigure}
\centerline{\psfig{figure=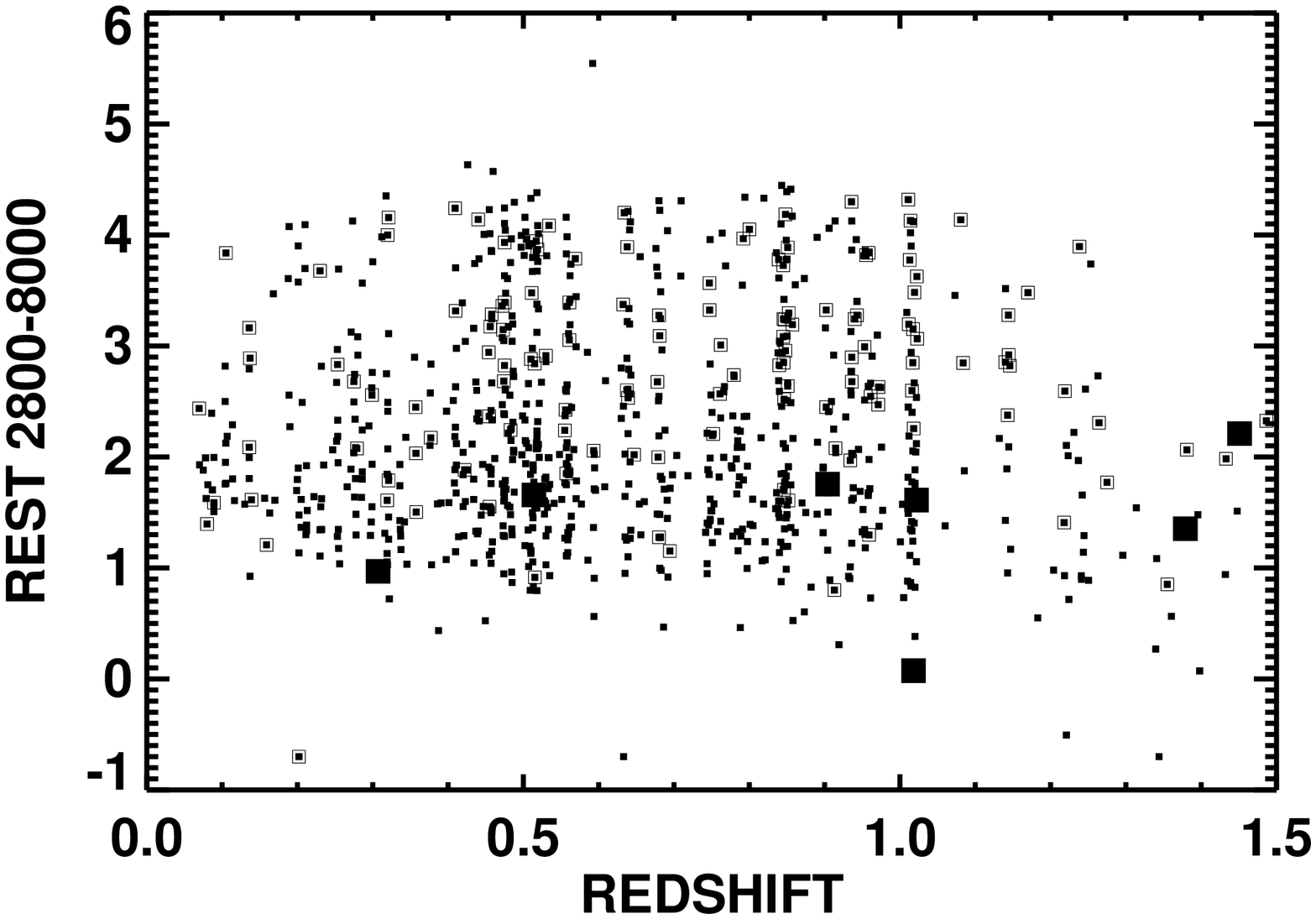,width=3.2in,angle=90}}
\caption{
Rest-frame $2800-8000$~\AA\ color versus redshift for the $z'<23.5$ 
sample. Only sources with spectroscopic redshifts are shown. Large 
solid squares denote broad-line AGNs, and large open squares denote 
{\it Chandra} sources.
\label{fig11}
}
\addtolength{\baselineskip}{10pt}
\end{inlinefigure}

In Figures~\ref{fig12}a and \ref{fig12}b, we also show the positions
of the sources with broad emission lines in their spectra
({\it large solid squares}) and the positions of the sources with high
X-ray luminosities ($L_{X}>10^{42}$~ergs~s~$^{-1}$; {\it large open squares}).
Most of the high ultraviolet luminosity sources fall in these categories
and may have substantial AGN contributions. Previous analyses have not
been able to distinguish between these sources and galaxies where the
ultraviolet light is dominated by star formation. Further discussion
of the evolution of the light density and luminosity functions with
redshift may be found in \markcite{capak04}Capak et al.\ (2004).

%
%
\begin{inlinefigure}
\centerline{\psfig{figure=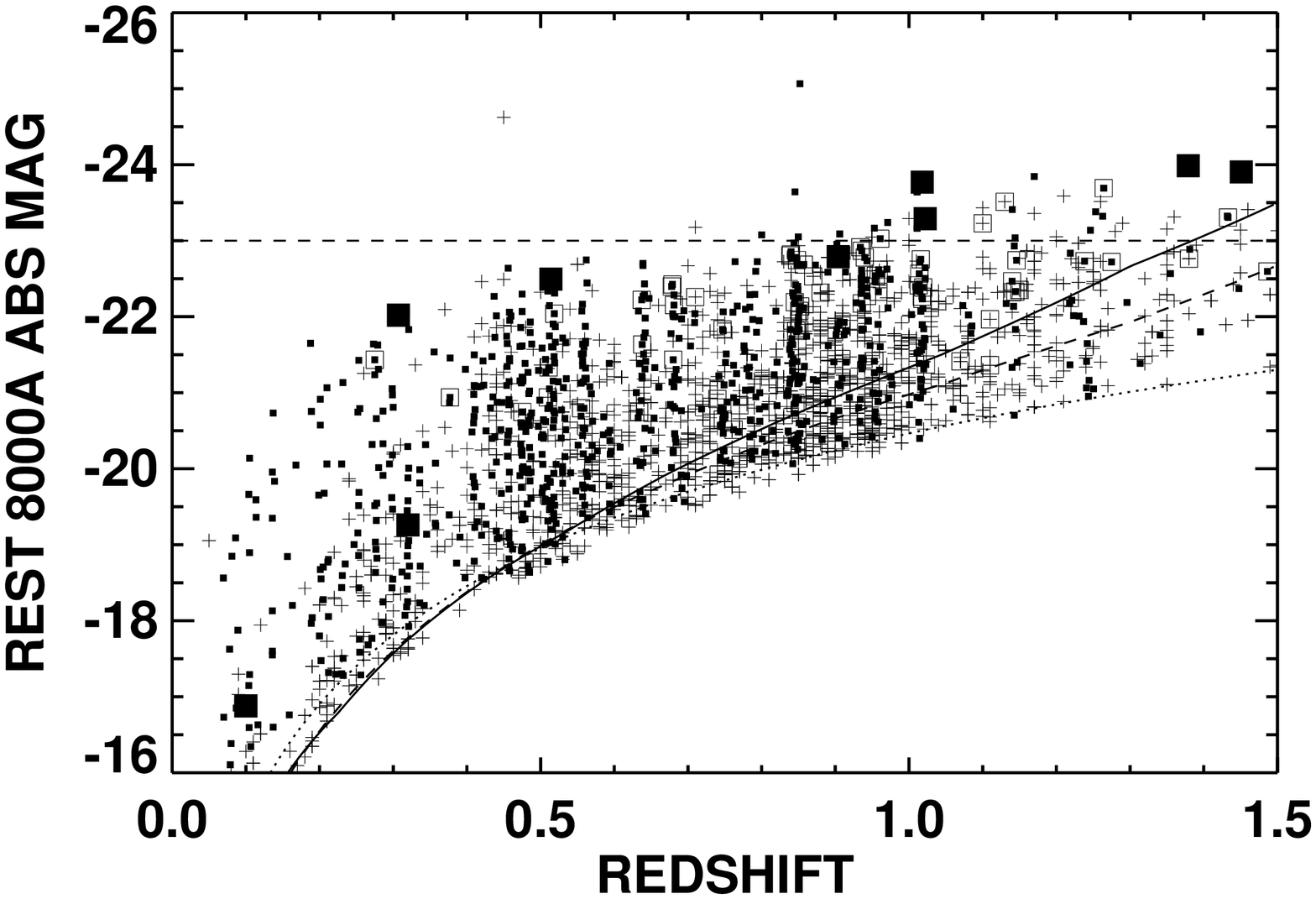,width=3.2in,angle=90}}
\centerline{\psfig{figure=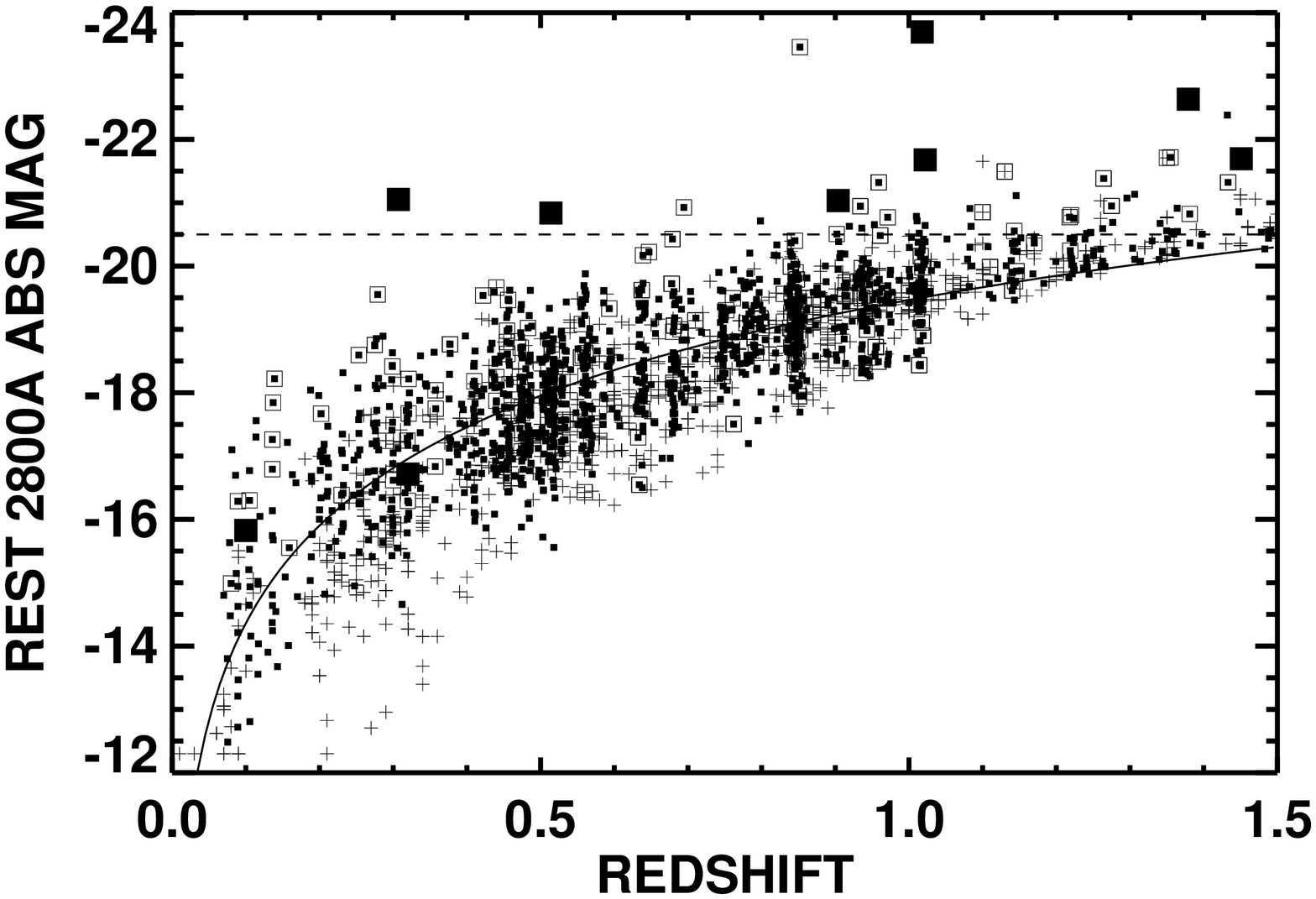,width=3.2in,angle=90}}
\caption{
Absolute rest-frame magnitude versus redshift at (a) 8000~\AA\
and (b) 2800~\AA. In each case, sources with spectroscopic redshifts 
are denoted by small solid symbols, and sources with photometric 
redshifts are denoted by plus signs. Sources with broad emission 
lines are denoted by large solid squares, and sources with 
X-ray luminosities $L_{X}>10^{42}$~ergs~s~$^{-1}$ by
large open squares. Dotted curve in (a) shows the $HK'<22.5$ selection 
for sources with a flat $f_{\nu}$ spectrum, while solid and dashed 
curves show the $z'<23.5$ selection for an elliptical galaxy and a 
spiral galaxy, respectively. Solid curve in (b) shows the $R<24$ 
selection for a flat $f_{\nu}$ spectrum galaxy.
The horizontal dashed lines show a reference absolute magnitude
of -23 for 8000~\AA and -21.5 for 2800~\AA which roughly
corresponds to the maximum observed values at high redshift
for objects without AGN signatures.
\label{fig12}
}
\addtolength{\baselineskip}{10pt}
\end{inlinefigure}

\subsection{X-ray properties versus environment}
\label{secx}

A large fraction of the most luminous galaxies harbor AGNs. 
For all galaxies with $M_{8000~{\rm \AA}}<-21$ in the $z=0-1$
range, based on either spectroscopic or photometric
redshifts, the fraction with $L_{X}>10^{42}$~ergs~s~$^{-1}$
is 6\%, while for $M_{8000~{\rm \AA}}<-22$, it is 15\%, in agreement
with previous estimates (e.g., \markcite{barger01}Barger et al.\ 2001).
Conversely, nearly all of the luminous X-ray sources
lie in optically luminous galaxies. Of the sources with
$L_{X}>10^{42}$~ergs~s~$^{-1}$, 82\% lie in galaxies with
$M_{8000~{\rm \AA}}<-21$ in the $z=0-1.2$ redshift range, where
lower optical luminosity sources would still be seen. 
Neither result shows any strong redshift dependence.

\markcite{barger02}Barger et al.\ (2002, 2003) and 
\markcite{gilli03}Gilli et al.\ (2003) noted that the X-ray 
sources are concentrated into velocity sheets. This has been taken 
to imply that the X-ray activity is found preferentially in higher 
density regions. However, it is not clear that the X-ray sources 
are more concentrated into these regions than the luminous field 
galaxies that contain them. 

%
%
\begin{inlinefigure}
\centerline{\psfig{figure=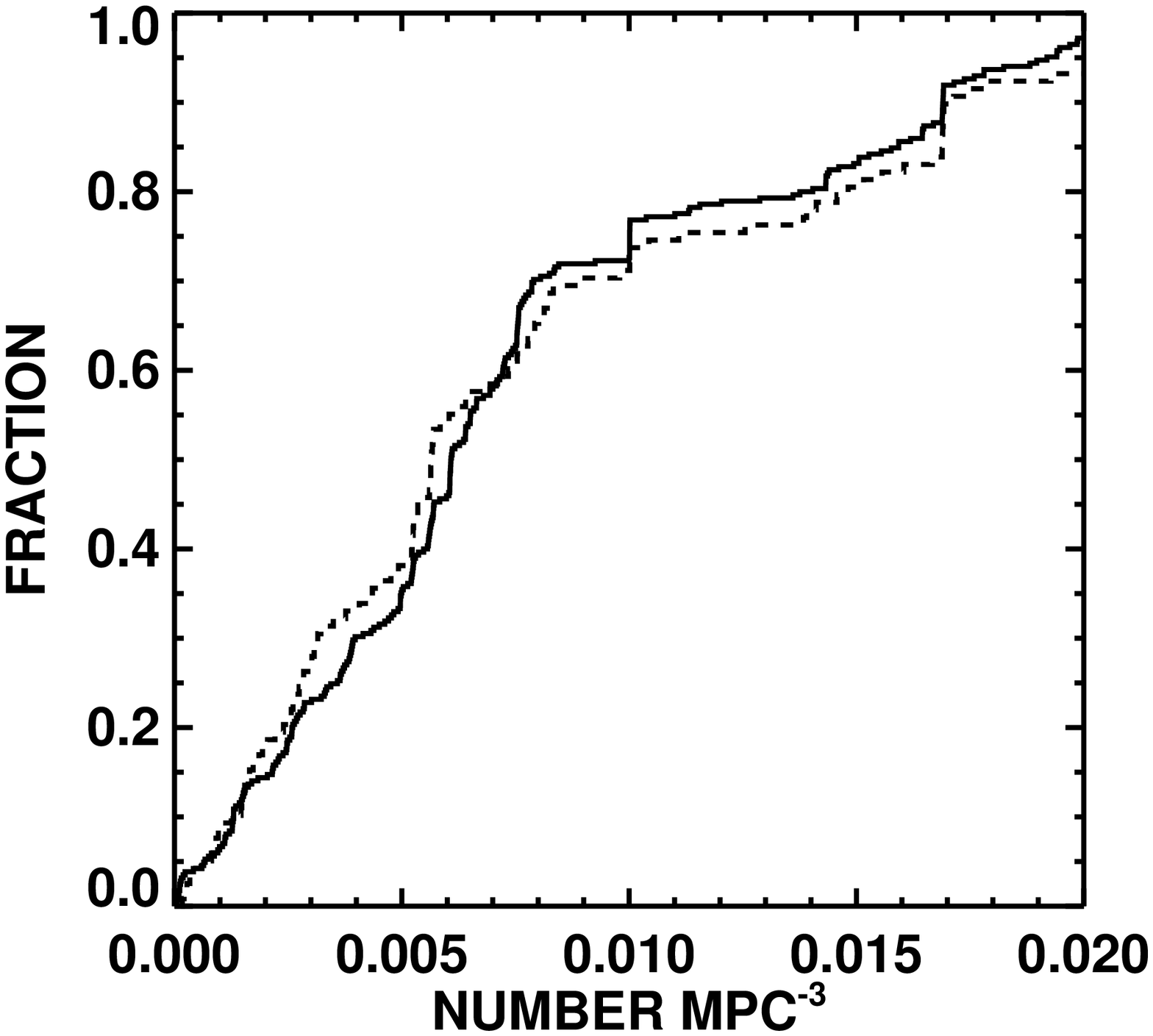,width=3.2in,angle=90}}
\caption{
Density environment of X-ray sources with
$M_{8000~{\rm \AA}}<-21$ and $L_{X}>10^{42}$~ergs~s~$^{-1}$ in
the redshift interval $z=0-1.2$ ({\it dashed line}) compared 
with that for field sources with $M_{8000~{\rm \AA}}<-21$ in the same 
redshift interval ({\it solid line}).  There is
no significant difference at the 95\% 
confidence level.
\label{fig13}
}
\addtolength{\baselineskip}{10pt}
\end{inlinefigure}

In order to test this, we estimated a simple neighborhood
density for each galaxy by measuring the number density
of $M_{8000~{\rm \AA}}<-21$ galaxies in a redshift interval of
$z=0.01$ about the source, averaged through the ACS-GOODS
rectangle. The density was computed using only the
spectroscopically identified sources and then corrected
to a total number density by multiplying by the ratio
of photometric to spectroscopic sources in the magnitude interval 
at that redshift. This density estimate is somewhwat crude since
it assumes the sheets are uniform structures over the
rectangle at all redshifts and also uses a redshift dependent
window to select the density, but it should be adequate for
the present purposes. We then used a Kolmogorov-Smirnov
test to compare the density environment of X-ray
sources with $M_{8000~{\rm \AA}}<-21$ and $L_{X}>10^{42}$~ergs~s~$^{-1}$
with field sources with $M_{8000~{\rm \AA}}<-21$. The two distributions
are shown in Figure~\ref{fig13}. The distribution of the X-ray
sources is
is not significantly different from that of the field sources
at the 95\% confidence level, and the two distributions
could be consistent. It appears that there
is no preference for the luminous X-ray sources to lie in higher
density environments.

\acknowledgements
Support was provided by NASA through grants HST-GO-09425.03-A (L.~L.~C.)
and HST-GO-09425.30-A (A.~J.~B.) from the Space Telescope Science 
Institute, which is operated by the Association of Universities for 
Research in Astronomy, Incorporated, under NASA contract NAS5-26555.
We also gratefully acknowledge support from NSF grants
AST-0084816 (L.~L.~C.), AST-0084847 (A.~J.~B.),
and AST-0071208 (E.~M.~H.), NSF CAREER award AST-0239425 (A.~J.~B.),
NASA grants DF1-2001X (L.~L.~C.) and GO2-3187B (L.~L.~C.),
the University of Wisconsin Research Committee with funds granted 
by the Wisconsin Alumni Research Foundation (A.~J.~B.), the 
Alfred P. Sloan Foundation (A.~J.~B.), and the David and
Lucile Packard Foundation (A.~J.~B.)


\pagestyle{empty}
\begin{deluxetable}{ccccccc}
\renewcommand\baselinestretch{1.0}
\tablewidth{0pt}
\tablecaption{Hawaii redshifts in the ACS-GOODS region}
\scriptsize
\tablehead{R.A. (J2000.0) & Decl. (J2000.0) & $z^\prime$ & $R$ & $B$ & $z_{spec}$}
\startdata
189.394943  &  62.24305  &  $ 18.5$  &  \nodata  &  $ 20.7$  &  star   & \cr
189.171753  &  62.12917  &  $ 16.6$  &  \nodata  &  $ 18.3$  &  star   & \cr
189.287094  &  62.29710  &  $ 17.5$  &  \nodata  &  $ 18.7$  &  star   & \cr
189.338516  &  62.21321  &  $ 18.7$  &  \nodata  &  $ 20.3$  &  0.104  & \cr
189.331253  &  62.21233  &  $ 17.8$  &  \nodata  &  $ 20.1$  &  star   & \cr
189.009216  &  62.26384  &  $ 17.7$  &  \nodata  &  $ 19.8$  &  0.087  & \cr
189.105804  &  62.23475  &  $ 16.8$  &  \nodata  &  $ 20.3$  &  star   & \cr
189.418152  &  62.36593  &  \nodata  &  \nodata  &  $ 19.1$  &  star   & \cr
\enddata
\label{tab1}
\end{deluxetable}
\pagestyle{empty}
\begin{deluxetable}{cccccrrrrrrrrll}
\renewcommand\baselinestretch{1.0}
\tablewidth{0pt}
\tablecaption{Spectroscopic redshifts from the Hawaii sample and the literature}
\scriptsize
\tablehead{\\ R.A. (J2000.0) & Decl. (J2000.0) & $HK^\prime$ & $z^\prime$ & $I$ & $R$ & $V$ & $B$ & $U$ & $z_{spec}$ & source & $z_{phot}$ & odds \\ (1) & (2) & (3) & (4) & (5) & (6) & (7) & (8) & (9) & (10) & (11) & (12) & (13)}
\startdata
189.394943  &  62.24305  &  $ 17.9$  &  $ 18.5$  &  $ 19.3$  &  \nodata  & 
 $ 19.6$  &  $ 20.7$  &  $ 21.8$  &  star   &   H  &  \nodata  &  \nodata
  \cr 
189.243301  &  62.15993  &  $ 17.0$  &  $ 16.7$  &  $ 17.4$  &  \nodata  & 
 $ 17.5$  &  $ 18.3$  &  $ 18.7$  &  star   &   C00  &  \nodata  &  \nodata
  \cr 
189.171753  &  62.12917  &  $ 16.6$  &  $ 16.6$  &  $ 17.3$  &  \nodata  & 
 $ 17.6$  &  $ 18.3$  &  $ 19.3$  &  star   &   H  &  \nodata  &  \nodata
  \cr 
189.287094  &  62.29710  &  $ 18.2$  &  $ 17.5$  &  $ 18.0$  &  \nodata  & 
 $ 18.0$  &  $ 18.7$  &  $ 18.5$  &  star   &   H  &  \nodata  &  \nodata
  \cr 
189.338516  &  62.21321  &  $ 18.2$  &  $ 18.7$  &  $ 18.8$  &  \nodata  & 
 $ 19.6$  &  $ 20.3$  &  $ 21.1$  &  0.1040  &   H  &  \nodata  &  \nodata
  \cr 
189.331253  &  62.21233  &  $ 17.4$  &  $ 17.8$  &  $ 18.6$  &  \nodata  & 
 $ 19.0$  &  $ 20.1$  &  $ 22.2$  &  star   &   H  &  \nodata  &  \nodata
  \cr 
189.009216  &  62.26384  &  $ 17.4$  &  $ 17.7$  &  $ 17.9$  &  \nodata  & 
 $ 18.9$  &  $ 19.8$  &  $ 20.9$  &  0.0870  &   H  &  \nodata  &  \nodata
  \cr 
189.105804  &  62.23475  &  $ 16.4$  &  $ 16.8$  &  $ 17.7$  &  \nodata  & 
 $ 19.1$  &  $ 20.3$  &  $ 22.2$  &  star   &   H  &  \nodata  &  \nodata
  \cr
\enddata
\label{tab1}
\end{deluxetable}
\pagestyle{empty}
\begin{landscape}
\begin{deluxetable}{ccccccccc}
\renewcommand\baselinestretch{1.0}
\tablewidth{0pt}
\tablecaption{Hawaii spectroscopic redshifts with discrepancies}
\scriptsize
\tablehead{\\ R.A. (J2000.0) & Decl. (J2000.0) & redshift & Deimos & LRIS & KTRS & C00 & D01 & comments \\ (1) & (2) & (3) & (4) & (5) &
 (6) & (7) & (8) & (9)}
\startdata
189.087875  &  62.20236  &  0.7794  &  0.7794  &  \nodata  &  \nodata  &
 0.8410  &  \nodata  &  redshift based on OII, Ca H and K, H beta and OIII  \cr
189.121078  &  62.23117  &   1.3980  &  \nodata  &  \nodata  &  0.4572  &
  1.3980  &  \nodata  &  OII line at 8945A  \cr
189.179794  &  62.18575  &   3.2340  &   3.2340  &  \nodata  &  0.2980  &
 0.2990  &  \nodata  &  Break interpreted as Lyman continuum  \cr
189.191132  &  62.22386  &  0.3200  &  \nodata  &  0.3200  &  0.3208  &
 0.3210  &  0.8470  &  Neither this object or object to W (z=0.4399) have 0.847  \cr
189.206039  &  62.22972  &  0.0890  &  0.8520  &  0.0890  &  0.0893  &  0.0890
  &  \nodata  &  Not clear where 0.852 redshift arises from. No sign of doubling  \cr
189.216583  &  62.22564  &   1.0870  &  \nodata  &  0.9590  &  \nodata  &
  1.0870  &  \nodata  &  relatively poor spectrum  \cr
189.220337  &  62.23472  &   1.0142  &  \nodata  &  0.4980  &   1.0142  &
 0.4980  &  \nodata  &  Hawaii redshift based on CaK and H beta in absorption is questionable  \cr
189.233917  &  62.22492  &   1.2420  &  \nodata  &   1.2420  &   1.2444  &
 0.2710  &  \nodata  &  Hawaii redshift is poor but OIII doublet identification looks real  \cr
\enddata
\label{tab1}
\end{deluxetable}
\end{landscape}
\end{document}